\documentclass[final,authoryear,5p,times,twocolumn]{elsarticle} %%, longtitle

\usepackage{graphics}
\usepackage{amssymb}
\usepackage{color}
\usepackage{fixltx2e}
\usepackage{aas_macros}
\journal{Astroparticle Physics}

\newcommand{\komm}[1]{#1}

\begin{document}

\begin{frontmatter}



\title{Performance of the MAGIC stereo system obtained with Crab Nebula data}


%\author{authors...}
%\address{addresses...}

% authors 22.7.2011  Format elsart
%
\author[a]{J.~Aleksi\'c}
\author[b]{E.~A.~Alvarez}
\author[c]{L.~A.~Antonelli}
\author[d]{P.~Antoranz}
\author[b]{M.~Asensio}
\author[e]{M.~Backes}
\author[b]{J.~A.~Barrio}
\author[f]{D.~Bastieri}
\author[g,h]{J.~Becerra Gonz\'alez}
\author[i]{W.~Bednarek}
% ...

\cortext[bbb]{Corresponding authors: J. Sitarek (jusi@kfd2.phys.uni.lodz.pl), E. Carmona (carmona@mppmu.mpg.de)}

\address[a]{IFAE, Edifici Cn., Campus UAB, E-08193 Bellaterra, Spain}
\address[b]{Universidad Complutense, E-28040 Madrid, Spain}
\address[c]{INAF National Institute for Astrophysics, I-00136 Rome, Italy}
\address[d]{Universit\`a  di Siena, and INFN Pisa, I-53100 Siena, Italy}
\address[e]{Technische Universit\"at Dortmund, D-44221 Dortmund, Germany}
\address[f]{Universit\`a di Padova and INFN, I-35131 Padova, Italy}
\address[g]{Inst. de Astrof\'{\i}sica de Canarias, E-38200 La Laguna, Tenerife, Spain}
\address[h]{Depto. de Astrof\'{\i}sica, Universidad de La Laguna, E-38206 La Laguna, Spain}
\address[i]{University of \L\'od\'z, PL-90236 Lodz, Poland}
% ...

\begin{abstract}
MAGIC is a system of two Imaging Atmospheric Cherenkov Telescopes located in the Canary island of La Palma. 
Since autumn 2009 both telescopes have been working together in stereoscopic mode, providing a significant improvement with respect to the previous single-telescope observations. 
We use observations of the Crab Nebula taken at low zenith angles to assess the performance of the MAGIC stereo system. 
The trigger threshold of the MAGIC telescopes is $50-60\,$GeV.
Advanced stereo analysis techniques allow MAGIC to achieve a sensitivity as good as $(0.76 \pm 0.03)$~\% of the Crab Nebula flux in 50$\,$h of observations above $290\,$GeV. 
The angular resolution at those energies is better than $\sim0.07^\circ$.
We also perform a detailed study of possible systematic effects which may influence the analysis of the data taken with the MAGIC telescopes. 

\end{abstract}


\end{frontmatter}

% \linenumbers

\section{Introduction}\label{intro}
MAGIC (Major Atmospheric Gamma Imaging Cherenkov telescopes) is a system of two 17~m diameter Imaging Atmospheric Cherenkov Telescopes (IACT). 
They are located at a height of 2200 m a.s.l. on the Roque de los Muchachos on the Canary Island of La Palma ($28^\circ$N, $18^\circ$W).
They are used for observations of particle showers produced by very high energy (VHE, $\gtrsim30\,$GeV) gamma-rays from galactic and extragalactic objects. 

While the first MAGIC telescope has been operating since 2004, the construction and the commissioning of the second telescope finished in autumn 2009.
Both telescopes are normally operated together in the so-called stereoscopic mode. 
In this mode, only events which trigger both telescopes are recorded and analyzed. 
Since the same event is seen by two telescopes, with a slightly different parallax angle, it allows for the full three-dimensional reconstruction of air showers.

The Crab Nebula is a nearby ($\sim2$~kpc away) pulsar wind nebula and the first source detected in VHE gamma-rays \citep{whipple_crab}.
Moreover, the Crab Nebula is the brightest steady VHE gamma-ray source, therefore it has become the so-called ``standard candle'' in VHE gamma-ray astronomy. 
Recent observation of flares in the GeV range (ATel \#2855, \citealp{ta11, ab10}) raised questions about the stability of the TeV flux. 
In fact a hint of increased flux was claimed by ARGO-YBJ (ATel \#2921), however it was not confirmed by the simultaneous observations of MAGIC and VERITAS (ATels \#2967, \#2968). 
In this paper we use the Crab Nebula data to evaluate the performance of the MAGIC telescopes. 
The scientific implications of the Crab Nebula observations and the stability of its flux will be addressed in a separate paper in preparation. 

In Section~\ref{sec:hard} we briefly describe the MAGIC telescopes. 
In Section~\ref{sec:data} we describe both, the Crab Nebula data sample, and the various Monte Carlo samples used in the analysis.
In Section~\ref{sec:analysis} we present the techniques and methods used for the analysis of the MAGIC stereo data.
In Section~\ref{sec:results} we evaluate the performance of the MAGIC stereo system. 
In Section~\ref{sec:systematics} we analyze and quantify various types of systematic errors. 
We conclude the paper with final remarks in Section~\ref{sec:concl}.

%-----------------------------------------------------------------------------
\section{The MAGIC telescopes}\label{sec:hard}
The second MAGIC telescope is a close copy of MAGIC~I \citep{magic_crab} with a few important differences. 
Both telescopes have a $17\,$m diameter mirror dish and $f/D$ close to 1, but contrary to MAGIC~I, the outer mirrors of MAGIC~II are made of glass rather than aluminium \citep{magic_mirrors}.
Due to the parabolic shape of the reflector, the time spread of synchronous light signals is very low ($< 1$~ns).
To correct for deformations of the dish and sagging of the camera during the observations, each mirror panel is equipped with 2 motors.
This system is referred to as the active mirror control (AMC).  
The light mechanical structure of both telescopes allows for a rapid repositioning time, necessary for observations of short phenomena such as gamma-ray bursts. 
The telescopes can perform a half-turn of $180^\circ$ in azimuth in just 20~s \citep{magic_drive}.

In order to be able to register short light pulses, each MAGIC telescope is equipped with a $3.5^\circ$ diameter camera with photomultipliers (PMTs) as individual pixels. 
In the MAGIC~I camera two types of pixels are used: the inner 397 PMTs have a diameter of 0.1$^\circ$, while the 180 outer ones are larger (0.2$^\circ$ diameter). 
The most central pixel of the MAGIC~I camera has been modified to be able to register also constant values of the flowing current. 
This pixel is normally used only for the observation of pulsars \citep{magic_centralpix}.
In contrast, the camera of MAGIC~II is homogeneous and consists of 1039 hexagonal pixels with a diameter of $0.1^\circ$ \citep{magic_camera}.
For the calibration of the PMT signals each telescope is equipped with a calibration box mounted in the middle of the reflector dish. 
The calibration box provides short calibration pulses of constant intensity uniformly illuminating the camera.
The calibration box of MAGIC~I is based on a set of LEDs in different colors \citep{magic_calib}, while the calibration box of MAGIC~II is composed of a frequency tripled Neodym YAG microchip laser and a set of attenuation filters. 

Electrical pulses at the output of the PMTs are converted into optical signals with the use of vertical cavity laser diodes (VCSELs) and transmitted over optical fibers into the counting house.
There, the signal from a given PMT is split in the so-called receiver boards into the trigger branch and the readout branch. 
The readout of the MAGIC~I telescope is based on multiplexed FADCs (Flash Analog to Digital converters).
The signals from groups of 16 pixels are artificially delayed by different time offsets and are plugged into a single FADC channel, sampling at 2GSamples/s \citep{magic_mux}.
Although the same sampling frequency is also used in MAGIC~II, its readout is based on the Domino Ring Sampler~2 (DRS2) chip, allowing for a lower cost and a more compact system.
The readout of the second telescope introduces a dead time of about 10\%, whereas the dead time of the MAGIC~I readout is negligible. 
Ultra-fast PMTs and readout electronics (together with the parabolic shape of the reflector) are the key elements in allowing us to use the time information in the reconstruction of the showers. 

For each pixel in the trigger region the Individual Pixel Rate Control (IPRC) software controls the discriminator threshold in real time (the so-called level-zero trigger). 
It keeps the accidental event rate coming from the night sky background (NSB) and the electronic noise at a constant level. 
Each telescope separately has a level-one trigger with the next neighbour (NN) topology.
$x$NN trigger topology means that only events with a compact group of at least $x$ pixels surviving the level-zero trigger pass the level-one trigger criterion. 
Unlike in mono observations, where 4NN topology is used, stereo data require a 3NN condition.
The trigger area for MAGIC~II is larger ($1.25^\circ$ radius) than for MAGIC~I ($0.95^\circ$ radius).
For special observations aiming for the lowest possible threshold, as an alternative to the level-one trigger, the so-called SUM-trigger is used \citep{magic_pulsar}.  
Both telescopes are working in stereo mode, i.e. only events triggering both telescopes are recorded. 
%%Since the stereo trigger system takes into account the geometric delay introduced by the position and the pointing of the telescopes, the width of the coincidence gate can be as short as  $100\,$ns.
\komm{
The level-one trigger signals of each telescope are propagated to the common stereo trigger board, where the coincidence is formed \citep{pa08}. 
Since the telescopes are separated by 85m the shower front will reach them at different times. 
Thus the level-one trigger signals are delayed by a value which is dependent on the pointing direction of the telescopes. 
With this procedure, the width of the coincidence gate is $100\,$ns.}

%-----------------------------------------------------------------------------
\section{Data sample}\label{sec:data}

In order to evaluate the performance of the MAGIC Stereo system, we use a sample of Crab Nebula data taken in November 2009 and January 2010. 
The data were taken at low zenith angles ($<30^\circ$).
After a data selection based on the rate of background events, 9$\,$h of good quality data were obtained. 
Observations were performed in the so-called wobble mode \citep{fo94}, i.e. with the source position offset by $0.4^\circ$ from the camera center in a given direction. 
This method allows us to simultaneously obtain the signal and background estimations at identical offsets from the viewing direction of the telescopes.
Every 20 minutes the direction of the wobble offset is inverted in order to decrease the systematic uncertainties induced by possible exposure inhomogeneities.
In addition, we analyze a few dedicated samples of Crab Nebula data taken at various offsets from the camera center. 
We use those samples to evaluate the performance of the MAGIC telescopes for sources at a non-standard distance from the camera center. 

For the analysis we use a sample of low zenith Monte Carlo (hereafter MC) gamma-rays generated with energies between 30$\,$GeV and 30$\,$TeV with the generation radius  $b_{\rm max}=350\,$m. 
In order to perform some basic background studies we also use diffuse MC samples of 
protons (30$\,$GeV -- 30$\,$TeV, simulated on a square $1.2\,$km $\times 1.2\,$km, viewcone semi-angle $\delta=5^\circ$), 
helium nuclei (70$\,$GeV -- 20$\,$TeV, $b_{\rm max}=1200\,$m, $\delta=6^\circ$ ) and 
electrons (70$\,$GeV - 7$\,$TeV, $b_{\rm max}=650\,$m, $\delta=4.5^\circ$).

%-----------------------------------------------------------------------------
\section{Analysis}\label{sec:analysis}

The analysis of the data has been performed using the standard software package called MARS (acronym for MAGIC Analysis and Reconstruction Software, see \citealp{magic_mars}).

\subsection{Image processing}
In the first steps of the analysis (calibration, image cleaning, and parametrization of images, see \citet{magic_crab, magic_time}) of the MAGIC stereo data each telescope is treated separately.
The procedure is similar for both telescopes with some small differences. 
The signal extraction in each channel of MAGIC~I uses a cubic spline algorithm. 
A number of FADC counts (after substraction of the pedestal) in time slices is interpolated by a third degree polynomial and then integrated \citep{magic_fadc}.
In MAGIC~II, the raw data has to be linearized before processing \citep{magic_daq}.
Afterwards the signal is extracted using a sliding window with a width of 4~ns from the total readout window of 40~ns.
The value of the baseline in the MAGIC~II channels is estimated from the first 8~ns of the same readout window \komm{on an event-by-event basis}.
The conversion from \komm{integrated} FADC counts to photoelectrons (phe) is done in both telescopes using the F-Factor (excess noise factor) method \citep[see e.g.][]{ml97}.
\komm{Due to different readout system in both telescopes, the typical conversion factors have much different value for MAGIC~II ($\sim0.01\,$phe/cnts) than for MAGIC~I ($\sim0.002\,$phe/cnts for inner pixels).}

\komm{The time response of the DRS2 chip is not homogeneous.
The channels of the Domino chip are subject to a time delay (up to a few ns) whose length depends on the absolute position of the readout window within the Domino ring.
Using the laser pulses taken during calibration runs we can calibrate this effect.}
%%%In addition, we perform a calibration of the time response of the DRS2 chip. 
This allows us to obtain a time resolution for large, simultaneous \komm{calibration} pulses as good as 0.33~ns\komm{, which is of he order of the sampling frequency}.

\begin{figure}[t!]
\centering 
\includegraphics[width=0.49\textwidth]{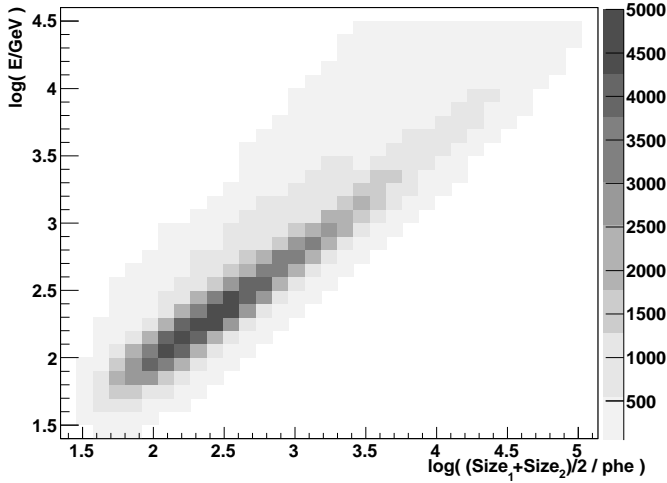}
\caption{
Correlation between the mean $Size$ of a gamma-ray event and its energy obtained with MC simulations. 
}\label{fig_energy_size}
\end{figure}

A typical air shower event produces a measurable signal in between a few and a few tens of pixels of the camera. 
The signal obtained in the rest of the pixels is induced by the NSB and the electronic noise.
The procedure used to select eventwise the pixels which come from a shower is called ``time image cleaning'' \citep{magic_time}. 
The time image cleaning used in MAGIC is a two stage process that comprises searching for \textit{core} and \textit{boundary} pixels. 
For MAGIC~I the \textit{core} of the image is composed of pixels with signals above 6$\,$phe.
Pixels with an arrival time different by more than 4.5~ns than the mean arrival time are excluded from the \textit{core}.
Single, isolated pixels are also removed from the image in order to avoid signals from PMT afterpulses.
An additional layer of \textit{boundary} pixels with a signal above 3$\,$phe and and an arrival time within 1.5~ns from the neighbouring \textit{core} pixel is added to the image in the second cleaning step.
For MAGIC~II, due to a higher light collection efficiency, a larger amount of pixels, and a somewhat higher noise, the thresholds are increased to 9$\,$phe and 4.5$\,$phe for \textit{core} and \textit{boundary} pixels respectively. 
The time constraints are the same for both telescopes.
Images are parametrized using the classical Hillas ellipses \citep{hi85}.
The total number of phe in the whole image, the $Size$ parameter, is strongly correlated with the energy of a gamma-ray event (see Fig.~\ref{fig_energy_size}).

Another cleaning algorithm, the so-called SUM image cleaning is implemented in MARS \citep{ri09}. 
In the case of SUM image cleaning groups of 2, 3 or 4 neighboring pixels with signals coming within a short time window are searched for. 
Each group is accepted if the sum of signals is above a given threshold (different for each multiplicity). 
Individual pixel signals are truncated before computing the sum to limit the effect of PMT afterpulses. 
In this paper we use the standard time image cleaning.
The SUM image cleaning will be discussed in another paper (in preparation). 

\subsection{Stereo parameters}

After the previous steps in the analysis chain, the images from both telescopes are combined to obtain the three-dimensional \textit{stereo} parameters: the $Impact$ parameters, defined as the orthogonal distances of the shower axes to the telescopes, and the height of the shower maximum (hereafter the $MaxHeight$).
The stereo parameters improve the energy estimation and the background rejection \citep{fe97}. 

\begin{figure*}[t!]
\centering 
\includegraphics[width=0.32\textwidth]{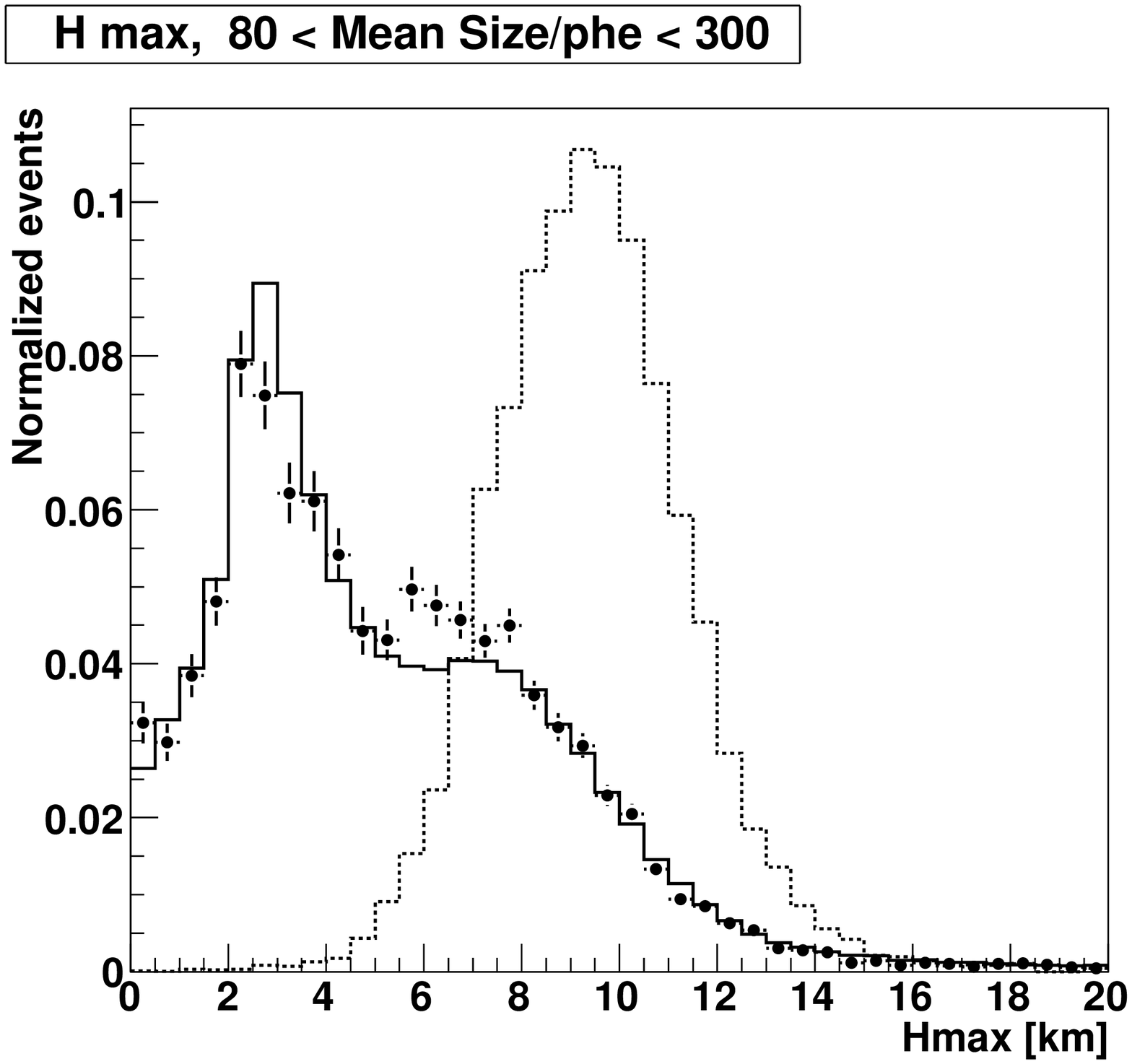}
\includegraphics[width=0.32\textwidth]{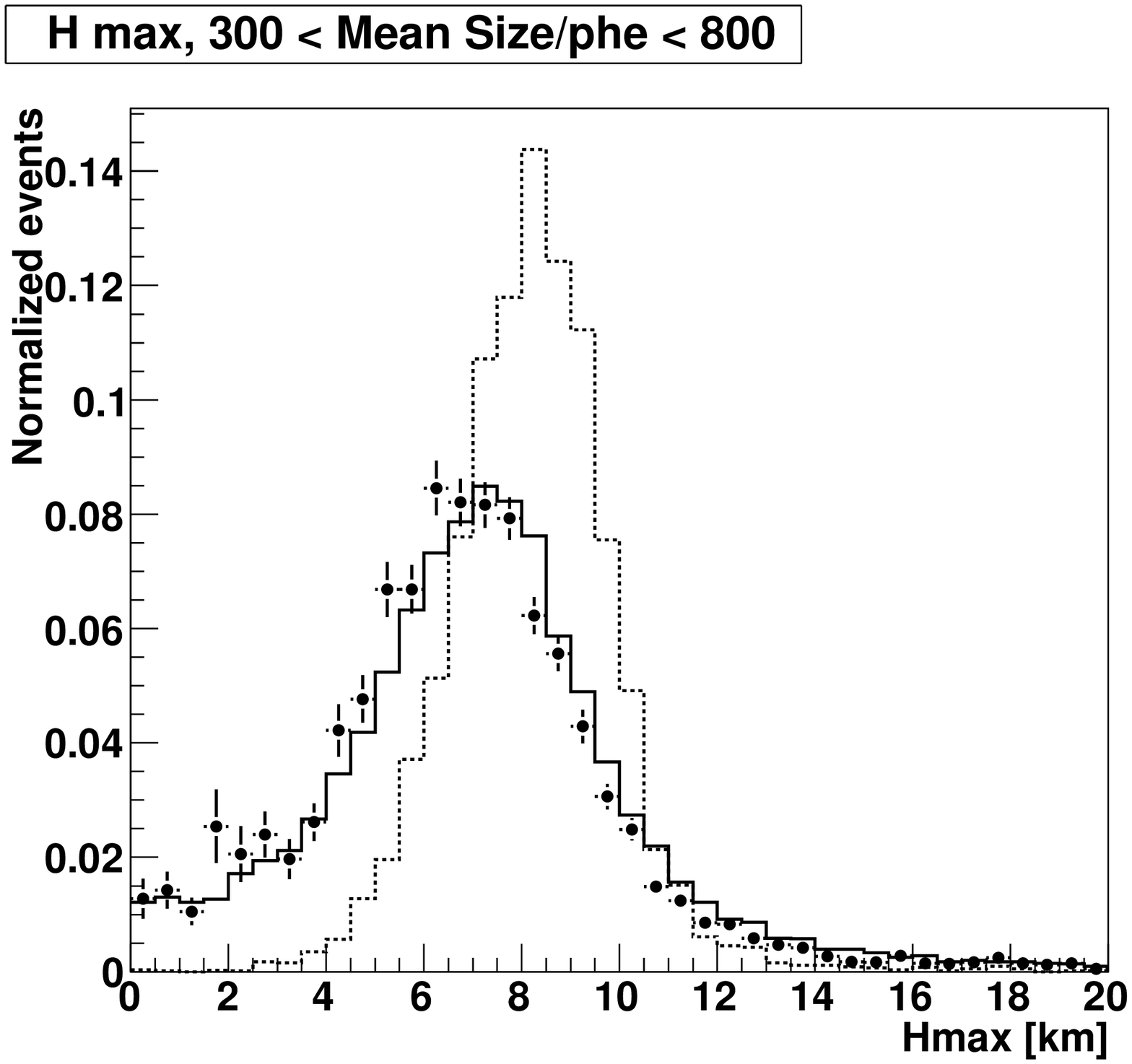}
\includegraphics[width=0.32\textwidth]{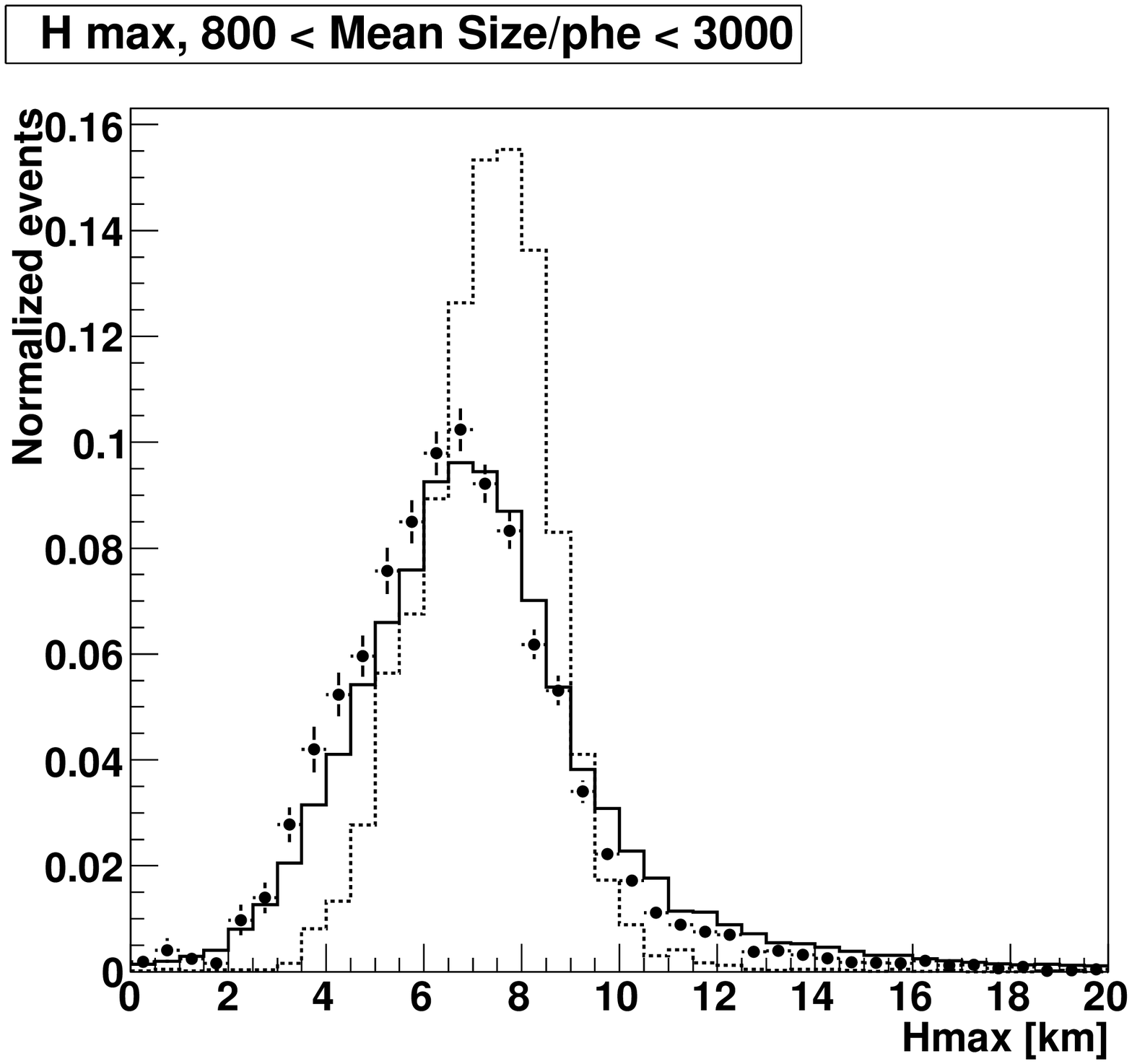}
\caption{
Distribution of $MaxHeight$ in different ranges of the mean $Size$ parameter for MC gamma-rays (dotted line), MC protons (full circles), and real data (solid line).
}\label{fig_hmax}
\end{figure*}

The distributions of $MaxHeight$ for different bins of image $Size$ (total number of phe in the image) are shown in Fig.~\ref{fig_hmax}.
The distribution of $MaxHeight$ for gamma-rays is a Gaussian. 
Higher energy showers penetrate deeper into the atmosphere.
Therefore, the maximum of the $MaxHeight$ distribution is shifted to lower values for larger $Size$. 
On the other hand, the shape of the $MaxHeight$ distribution for background events is more complicated.
For large images ($Size>300\,$phe), hadronic events have a single-peaked distribution of $MaxHeight$, which is a bit broader than for gamma-ray showers.
In contrast, for events with $Size<300\,$phe a second peak appears at the height of $\sim 2-3\,$km above the telescopes.
This additional peak is produced by single muon events \komm{(see also \citealp{mk07})}.
Therefore, $MaxHeight$ is a powerful parameter, when used for gamma/hadron separation at low energies. 
It is especially important taking into account that small images usually have poorly determined Hillas parameters.

\subsection{Gamma/hadron separation}
For the gamma/hadron separation in the stereo observations we use the same method, the so-called Random Forest (RF), as has been used for the single-telescope analysis \citep{magic_rf}.
The RF combines individual parameters from each telescope with the stereo parameters to produce a single number, the so-called $Hadronness$, which characterizes the likelihood of a given event to be of hadronic origin. 
In total, 11 parameters are used for the gamma/hadron separation.
In addition to the standard Hillas parameters of each telescope ($\log(Size_1)$, $\log(Size_2)$, $Width_1$, $Width_2$, $Length_1$, $Length_2$) and the stereo parameters ($Impact_1$, $Impact_2$, $MaxHeight$) we also use the gradients of the arrival times of the \komm{signals in the} pixels projected to the main axis of the image ($TimeGradient_1$, $TimeGradient_2$). 
Those parameters do not assume any a priori known source position.
Therefore, the obtained $Hadronness$ is not biased by the location of the source and can also be used to produce a skymap of the observed region.

\begin{figure}[t]
\centering 
\includegraphics[width=0.49\textwidth]{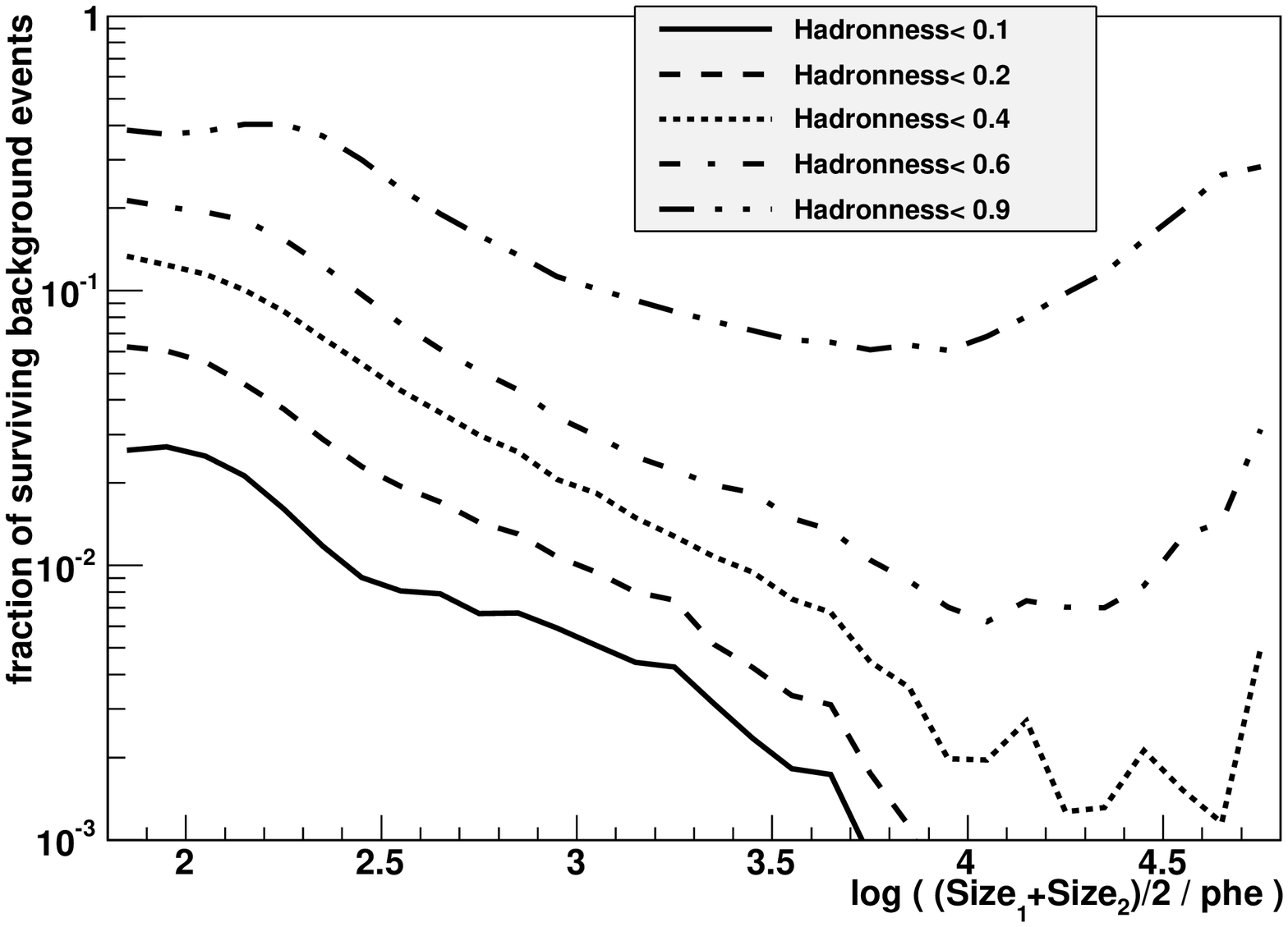}\\
\includegraphics[width=0.49\textwidth]{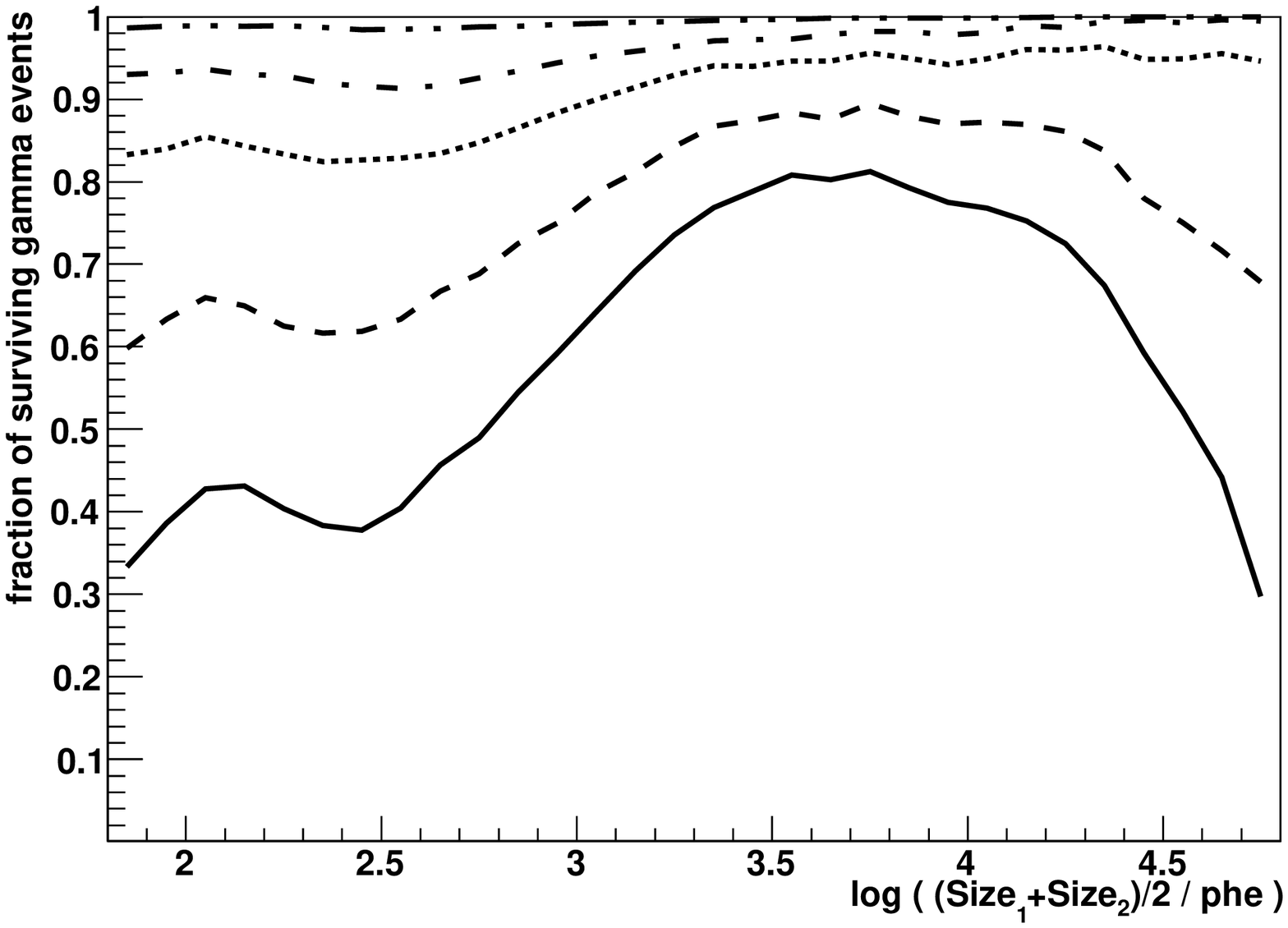}
\caption{Fraction of surviving background (data, top panel) and gamma-ray (MC, bottom panel) events. 
Different line styles are for different cuts: $Hadronness<0.1$ (solid), $<0.2$ (dashed), $<0.4$ (dotted), $<0.6$ (dot-dashed), and $<0.9$ (dot-dot-dashed).
}\label{fig_hadr}
\end{figure}

\begin{figure}[t]
\centering 
\includegraphics[width=0.49\textwidth]{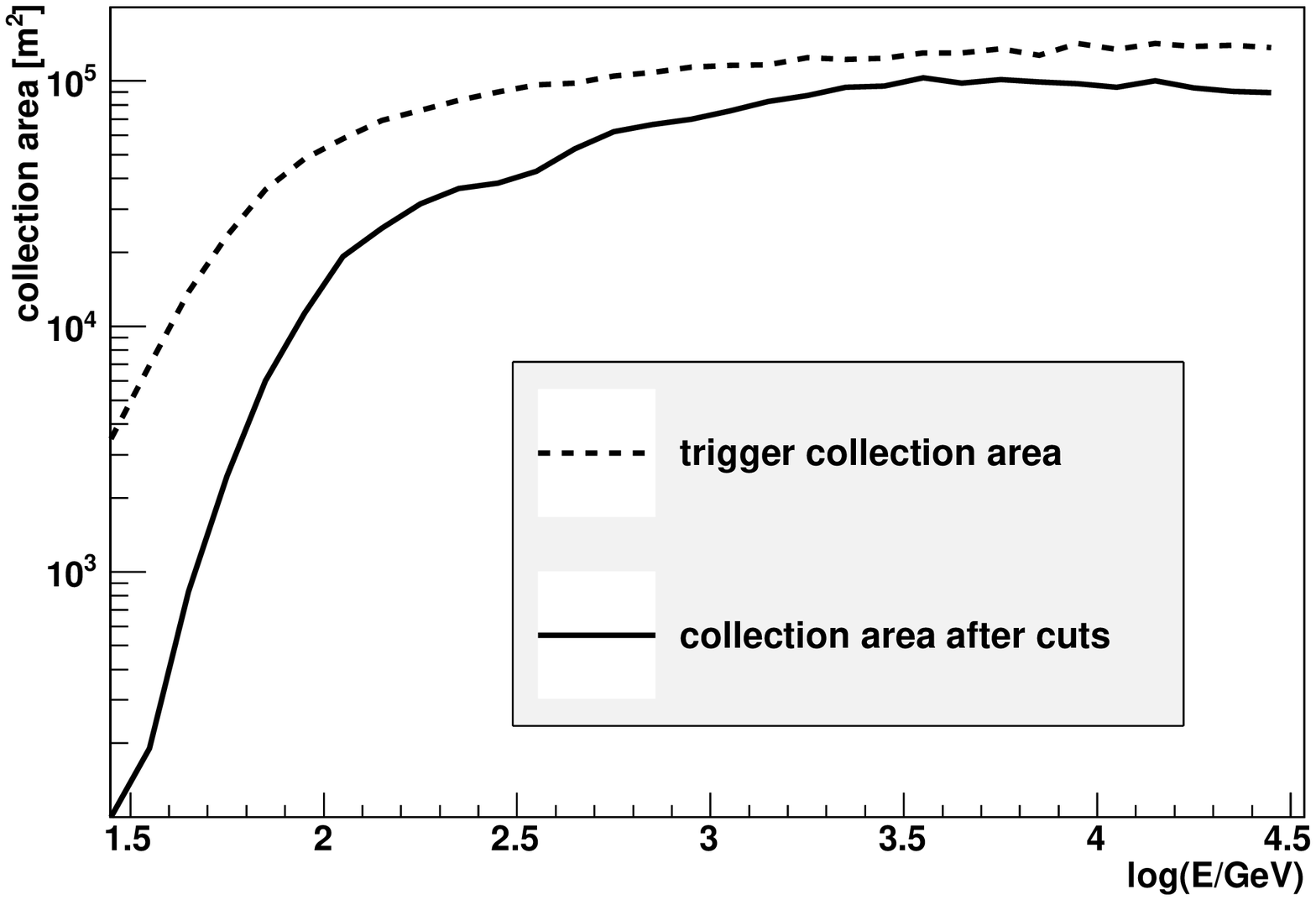}
\caption{
Gamma-ray collection area of the MAGIC telescopes (MC simulations) at the trigger level 
(dashed line) and after all analysis cuts (image cleaning of both images, $Size>50$ in both telescopes, gamma/hadron and $\theta^2$ cuts).
}\label{fig_collarea}
\end{figure}

The survival probability for gamma-rays and background events is shown in Fig.~\ref{fig_hadr} for different cuts in the $Hadronness$ parameter. 
At medium and higher energies, the $Hadronness$ allows us to achieve a reduction of the background by a factor of $\sim100$ with only a mild loss of gamma-ray events. 
The bump and dip structure clearly visible in the $Hadronness$ of gamma-rays is connected with the internal transition of the separation mostly based on the $MaxHeight$ parameter (at lower energies) to the one based on the Hillas parameters. 
The separation power drops at very high energies, due to a high fraction of truncated events and small statistics in the training sample. 
%The collection area of the MAGIC telescopes for the cuts optimized energy-wise for best differential sensitivity (see Section~\ref{sec_diffsens}) is presented in Fig.~\ref{fig_collarea}.
The collection area of the MAGIC telescopes \komm{using cuts optimized as a function of energy, so as to provide the best differential sensitivity} (see Section~\ref{sec_diffsens}) is presented in Fig.~\ref{fig_collarea}.
Because of the threshold effect, the collection area after all analysis cuts is rather steep \komm{function of energy} below 100$\,$GeV and \komm{reaches} a plateau at TeV energies.

\subsection{Energy reconstruction}
The energy of an event is estimated by averaging individual energy estimators for both telescopes. 
These estimators are derived from a look-up table based on a simple model for the light distribution of a shower based on the $Size$, $Impact$ \komm{parameters of one telescope plus the $MaxHeight$ parameter and} the zenith angle of the observation. 
Assuming that most of the light produced by a gamma ray in the atmosphere is contained in a light pool of radius $r_c$, the mean photon density in the light pool from a single charged particle of the gamma-ray shower can be calculated from \komm{the} total power of emitted light by such a particle at a given height in the atmosphere. 
Given the energy, $MaxHeight$ and zenith angle of a gamma-ray event, the amount of light produced, the light pool radius ($r_c$) and the mean photon density in the pool from a single particle ($\rho_c$) are computed using a simple atmospheric model. 
We use look-up tables filled with MC information of ($E_{true} \times \rho_c / Size$) as a function of $Impact/r_c$ and $Size$ for each telescope, to obtain the energy of an event measured by each telescope. 
This parametrization takes advantage of the fact that $E_{true}$ is roughly proportional to the number of secondary particles in the shower maximum, which scales as $Size/\rho_c$ and that the zenith angle dependence is automatically included. 
Using the MC tables generated in the previous step, the energy is calculated for each telescope from the values of $Impact$, $r_c$, $\rho_c$ and $Size$ computed in each event. 
The final energy estimation, $E_{rec}$, is the average value obtained from both telescopes (weighted with the inverse of the error).

\subsection{Event arrival direction reconstruction}

Classically in the stereo systems of Cherenkov telescopes the arrival direction of the primary particle is estimated as the crossing point of the main axes of the Hillas ellipses in the individual cameras \citep{ah97, ho99}. 
This method uses only geometrical properties of the images, neglecting the timing information. 

\komm{As was} shown for the case of the MAGIC~I stand-alone telescope \citep{magic_halo}, the inclusion of timing information can considerably improve the angular resolution, and as a result also the sensitivity. 
In that analysis the so-called DISP parameter (the angular distance between center of gravity of the image and the estimated source position) was estimated using multidimensional decision trees (the so-called DISP~RF method).
Then for each image, the source position was reconstructed to be DISP distant from the centre of gravity along the main axis of the image.

\begin{figure}[t]
\centering 
\includegraphics[scale=1., angle=90, trim=70 0 0 0, clip]{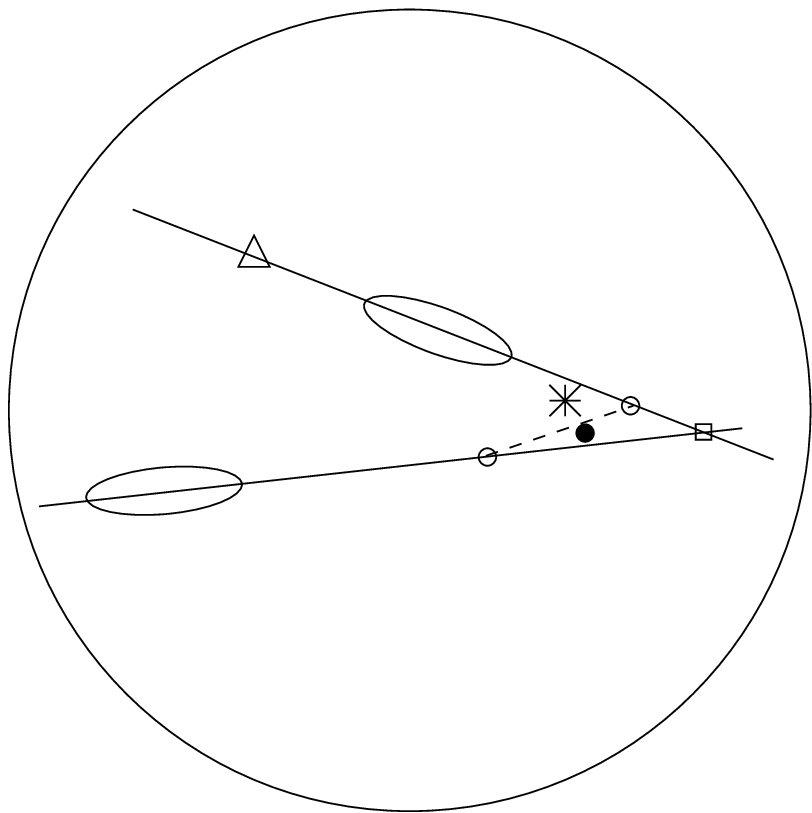}
\caption{
Principle of the Stereo DISP RF method.
The crossing point of the main axes of the images is shown as an empty square, and the DISP RF reconstructed position from each telescope is an empty circle.
The final reconstructed position (full circle) is a weighted average of those three points. 
The true source position is shown with a star.
\komm{Open triangle shows the reconstructed source position for the top image if the head-tail discrimination failed.}
}\label{fig_disprf}
\end{figure}

For stereo observations we apply the modified DISP RF method (hereafter Stereo DISP RF).
First, in addition to the crossing point of the main axes, we calculate also the DISP value for images in each telescope separately.
If the DISP positions from both telescopes agree with each other within a given radius we compute a weighted average of them (see Fig.~\ref{fig_disprf}). 
%
%In the opposite case, we check which individual DISP position is close to the crossing point of the main axes.
\komm{However it may happen, that the direction of DISP is misreconstructed and the source position is located on the other side of the image (wrong head-tail discrimination).
Therefore, if the reconstructed DISP positions from both telescopes are far apart, we check if one of the possible DISP position is close to the crossing point of the main axes.
In the opposite case (i.e. none of the DISP positions agrees with the crossing point)} the event is rejected.
In this way the angular resolution is always obtained using the most efficient method.

%%Recently, a simplified version of the above procedure, choosing the closest match among the two reconstructed directions per telescope, was found to deliver a similar performance and is now used in MAGIC as the default algorithm.
Recently, a simplified version of the above procedure, choosing the closest match among the \komm{four possible combinations of the two DISP positions in each telescope}, was found to deliver a similar performance and is now used in MAGIC as the default algorithm.
That method does not require the crossing point information.

\begin{figure}[t]
\centering 
\includegraphics[width=0.49\textwidth]{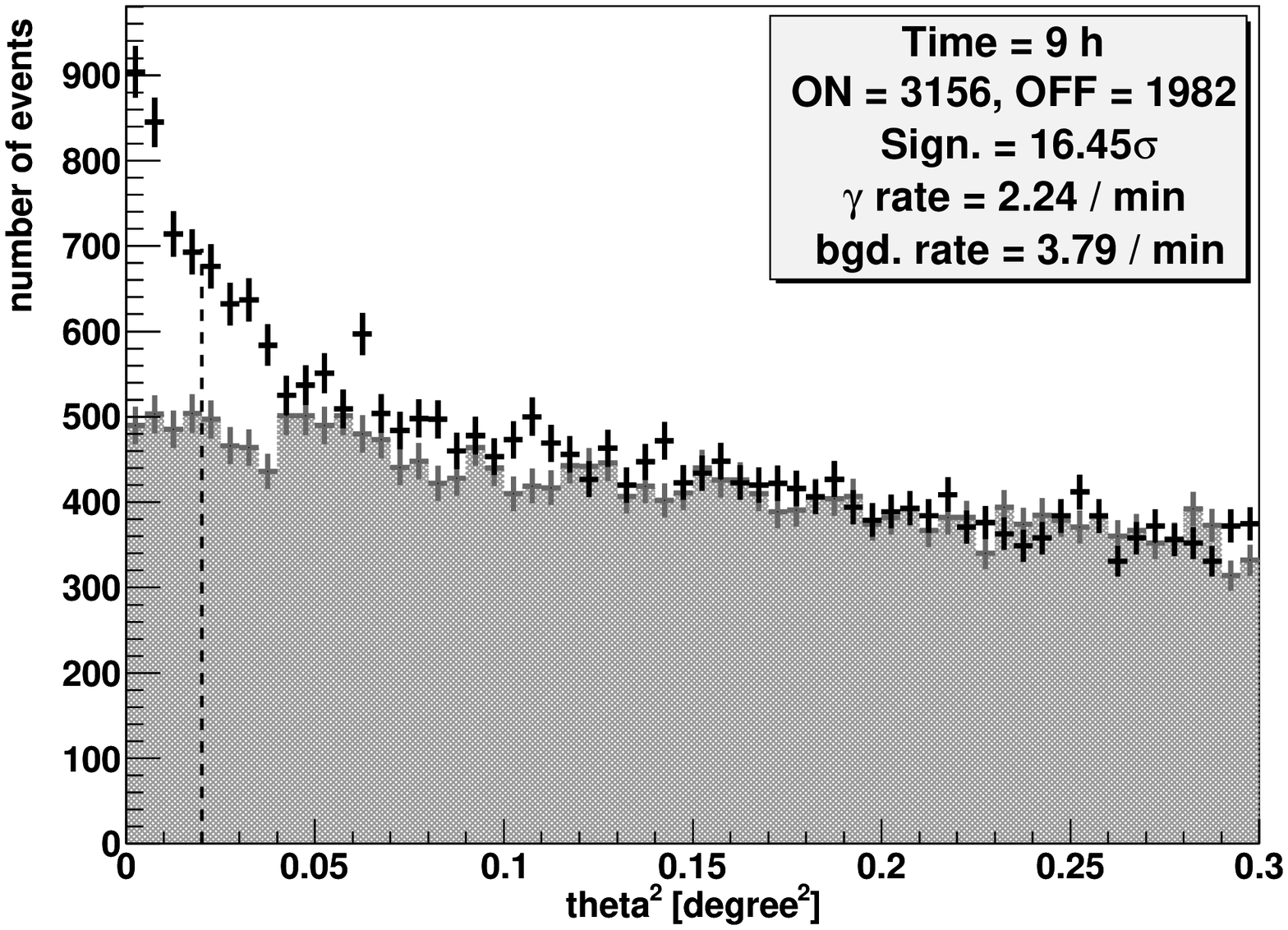}\\
\includegraphics[width=0.49\textwidth]{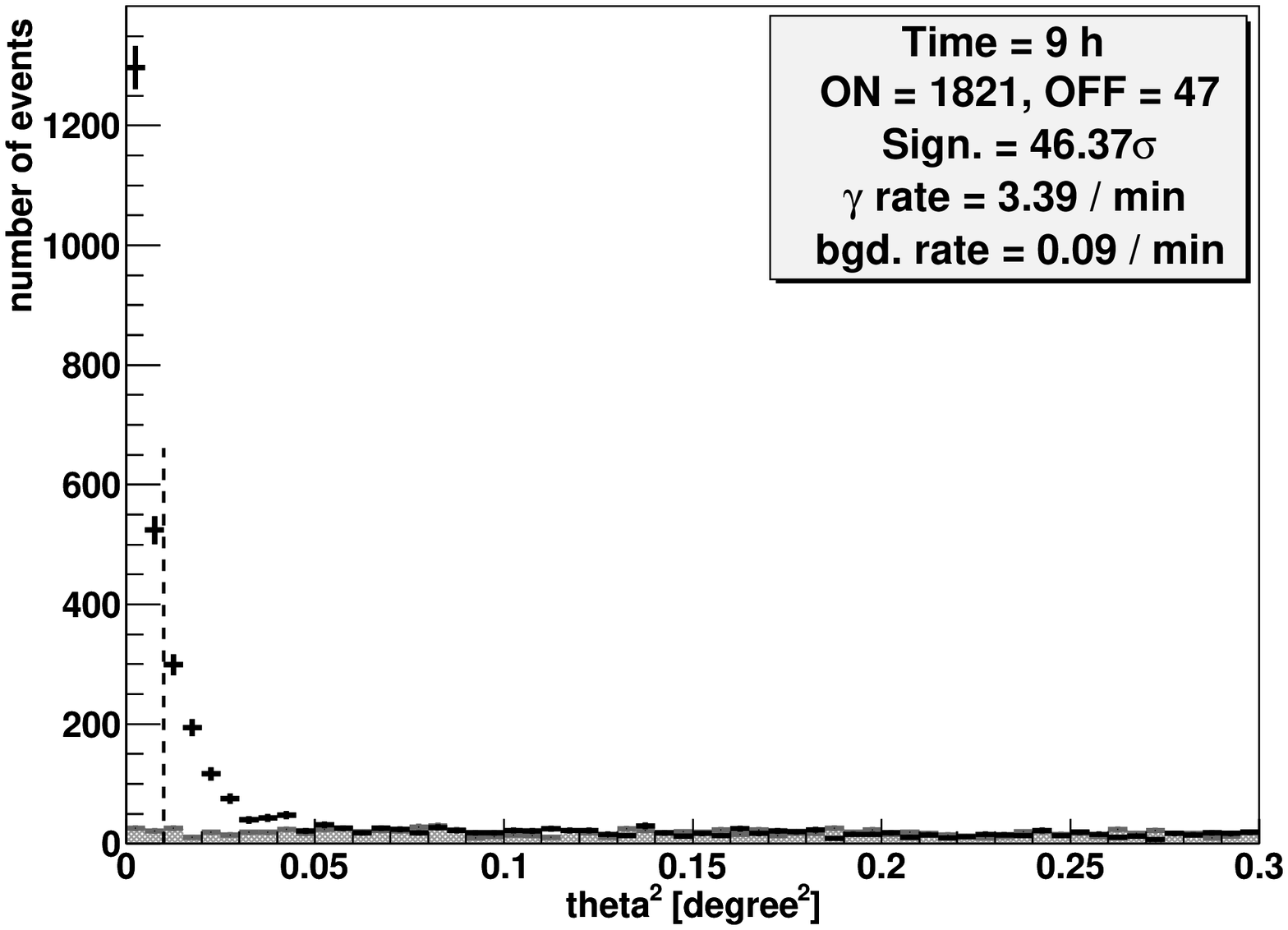}
\caption{
Theta$^2$ distribution of the signal (black points) in the analyzed sample of the Crab Nebula data and the background estimation (grey shaded area). 
Estimated energy $60\mathrm{GeV}<E_{est}<100\mathrm{GeV}$ (top panel) and 
$E_{est}>300\mathrm{GeV}$ (bottom panel). 
The cut in theta$^2$ used for the calculation of the number of ON and OFF events is shown with the dashed, vertical lines. 
}\label{fig_theta2}
\end{figure}

In addition, since the DISP method is trained on gamma-rays, often both reconstructed DISP positions do not agree for hadrons, and the event is rejected.
Therefore, an additional gamma/hadron separation factor is achieved at this step, which depending on the strength of the hadronness cut can remove between 10\% and 50\% of the background events.
The so-called $\theta^2$ plots (the distribution of the squared distance between the true and the reconstructed source position) in two energy ranges are shown in Fig.~\ref{fig_theta2}.
Observations with the MAGIC telescopes yield a highly significant signal below 100$\,$GeV. 
Moreover, a strong source such as the Crab Nebula allows for a nearly background free analysis above 300$\,$GeV.

%-----------------------------------------------------------------------------
\section{Results}\label{sec:results}

\subsection{Trigger threshold}
In order to achieve the lowest energy threshold, the  level-zero trigger amplitudes for the individual pixels are chosen as low as possible while keeping a negligible accidental stereo trigger rate.
As a result, both telescopes have a level-one trigger rate dominated by NSB triggers and PMT afterpulses.
Typical level-one trigger rates are around 10$\,$kHz for MAGIC~I and around 15$\,$kHz for MAGIC~II, due to the larger trigger area of MAGIC~II.
The resulting accidental stereo trigger rate is just $\sim 10\,$Hz, well below the stereo event rate from cosmic rays which typically is within the range 150--200$\,$Hz.
Accidental events are removed in the image cleaning procedure.

The energy threshold of the stereo system is estimated from Monte Carlo simulations where the individual pixel thresholds match those used during data taking. 
Nominal values of the individual pixel thresholds should be around 4$\,$phe per $0.1^\circ$ pixel, but the values are automatically adjusted during data taking to keep the accidental trigger rate at a low level. 
To avoid a bias in the conversion from discriminator threshold (amplitude measured in mV) to photoelectrons (which is an integrated charge), that will translate in further systematic errors, the values of the discriminator thresholds are estimated from the data. 
This is done by studying the pixel amplitude of the smallest showers that could trigger the system. 
The charge distribution of the pixel with the lowest charge in the 3NN pixel combination with largest total charge (the charge of the 3 pixels added) is used to obtain the value of the discriminator threshold measured in phe. 
For single telescope triggers, the peak of the distribution should be very close to the actual number of photoelectrons needed to trigger a pixel. 
For the stereo system it is biased to higher values and a comparison with MC simulations for different pixel threshold settings is needed in order to find the best match with the data distribution.

\begin{figure}[t]
\centering 
\includegraphics[width=0.49\textwidth]{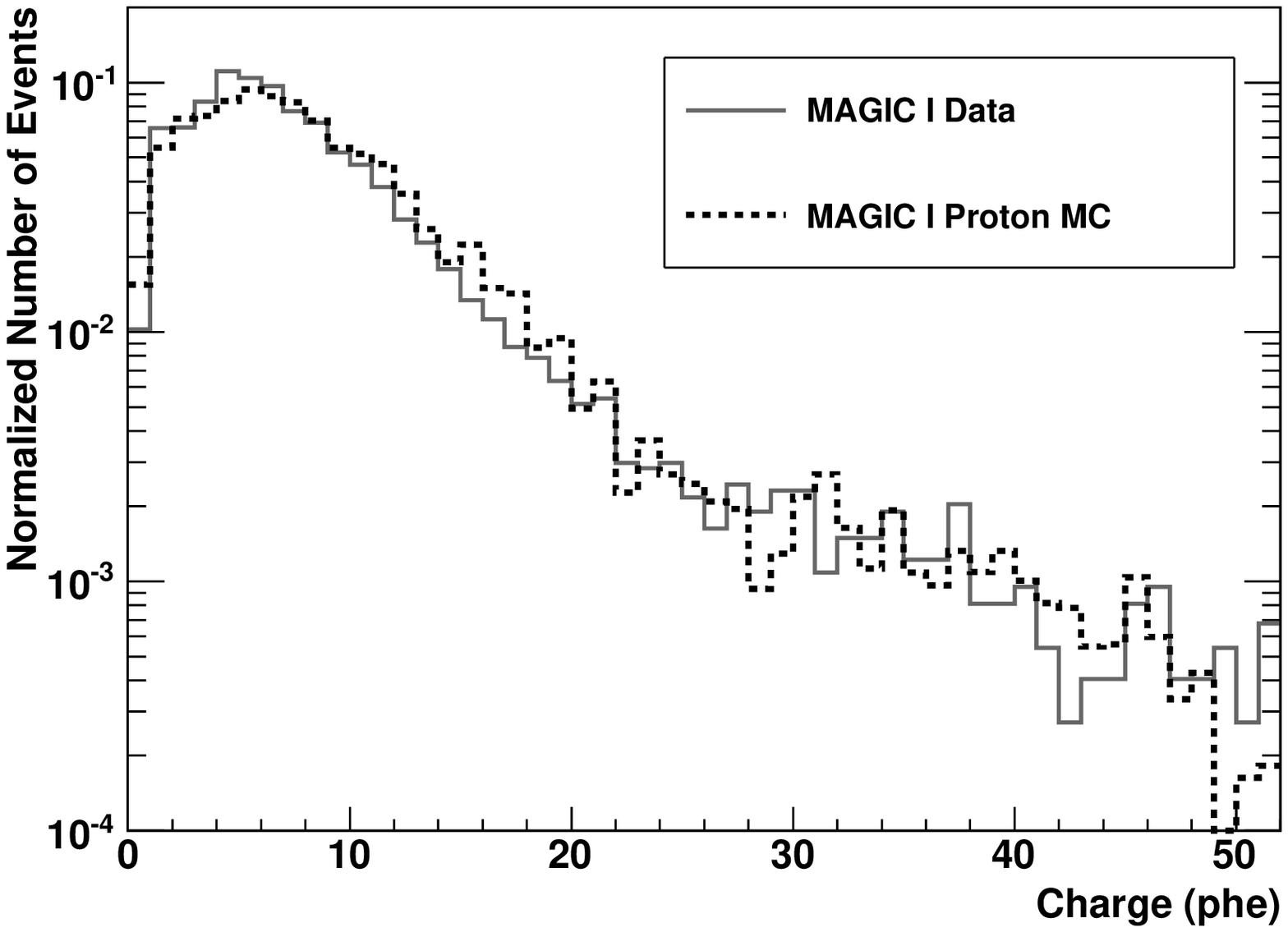}\\
\includegraphics[width=0.49\textwidth]{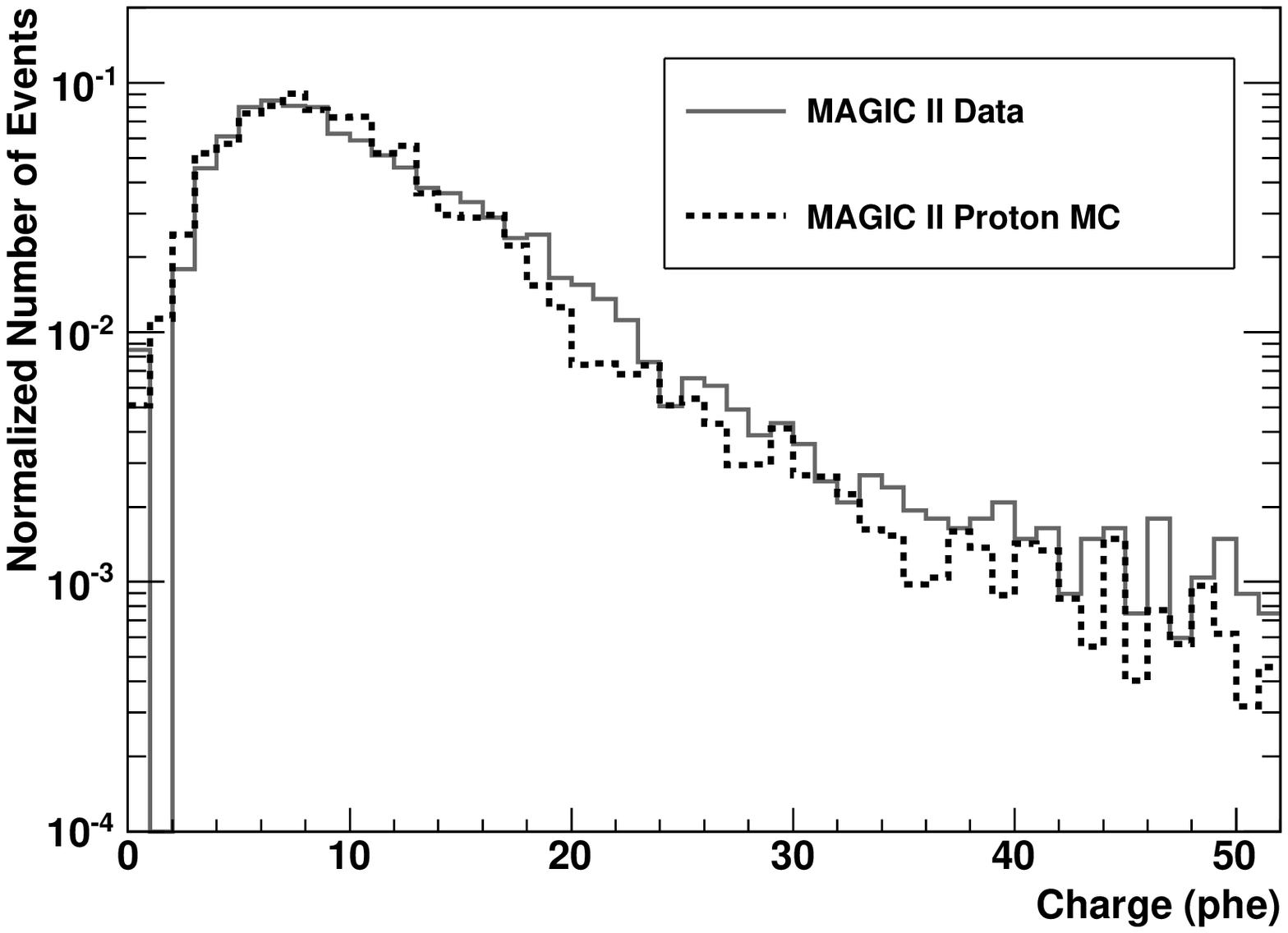}
\caption{
Distribution of the smallest charge of the brightest 3NN cluster for data and Monte Carlo (continuous and dotted lines respectively) for MAGIC~I (top panel) and MAGIC~II (bottom).
}\label{fig_thresholds}
\end{figure}

The distribution of the minimum single pixel charge in the 3NN combinations computed from data taken from a galactic source observed at a low zenith angle are shown in Fig.~\ref{fig_thresholds} for MAGIC~I and MAGIC~II. 
The MC distributions also shown in the same figure are those that match best the measured distribution on the data. 
The value of the pixel threshold is obtained from the values used in the MC simulation shown in the same figure. 
We obtain that the average threshold is 4.3$\,$phe in MAGIC~I and 5.0$\,$phe in MAGIC II. 
From the MC simulations we also obtain that the energy threshold, defined as the maximum of the distribution of triggered gamma-ray events for sources with a differential spectral index -2.6 is 50 GeV. 
If we use instead the measured, curved Crab Nebula spectrum, the maximum of the distribution of triggered gamma-ray events is 60$\,$GeV. 
After applying the image cleaning and cuts used in this analysis, the energy threshold increases to 75$\,$GeV.

\subsection{Relative light scale between both telescopes}
The mean amount of Cherenkov light on the ground produced by a VHE gamma-ray shower depends mostly on its energy and the impact parameter (note however also the dependence on the relative position with respect to the magnetic North due to the geomagnetic field effect, mostly pronounced at lowest energies, see e.g. \citealp{magic_gf}). 
Therefore, it is possible to investigate the relative light scale of both telescopes by selecting gamma-like events with a similar reconstructed impact parameter in both telescopes \komm{\citep{hof03}}. 
We apply rather strict cuts $Hadronness<0.2$ and $\theta^2<0.01$ to obtain a pure gamma-ray sample for this test. 
Hadronic background events would spoil the resolution of this method due to the strong internal fluctuations and poor estimation of the impact parameter. 
\begin{figure}[t]
\centering 
\includegraphics[width=0.49\textwidth]{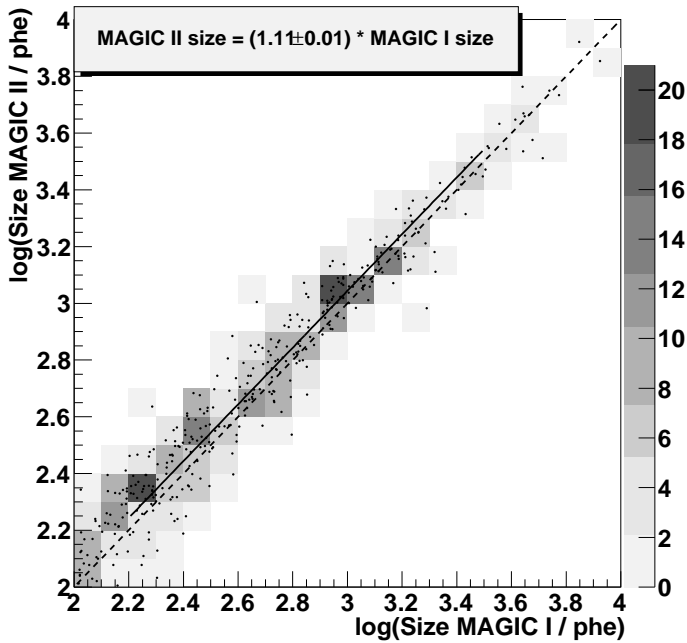}
\caption{
Relation between $Size_2$ and $Size_1$ for gamma-ray events obtained with a Crab Nebula sample. 
Only events with a difference in the reconstructed impacts of both telescopes smaller than $10$~m are considered. 
Individual events with $Hadronness<0.2$, $\theta^2<0.01$ and $Impact$ parameters $<150$~m are shown by black dots.
The black, solid line shows the result of the fit, the dashed line corresponds to $Size_1=Size_2$.
}\label{fig_sizes}
\end{figure}
We obtain that the absolute light scale for MAGIC II is $\sim$ 10\% larger than that of MAGIC~I \komm{(see Fig.~\ref{fig_sizes})}. 
This difference is within the systematic error of the absolute energy scale (see discussion in Section~\ref{sec:systematics}).

\subsection{Comparison of parameters}\label{sec:parameters}

\begin{figure}[t]
\centering 
\includegraphics[width=0.49\textwidth]{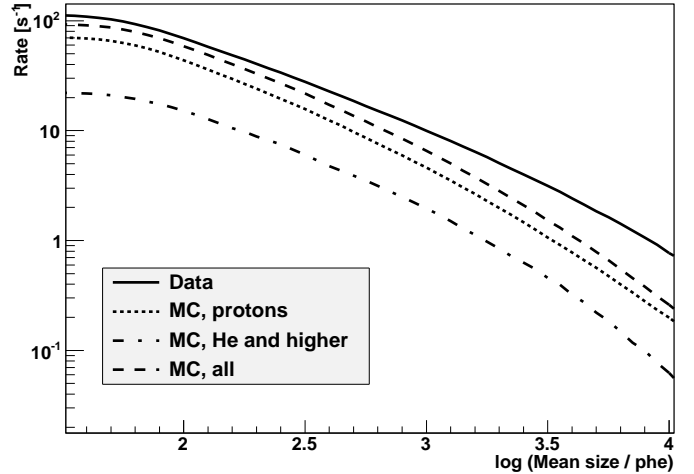}
\caption{
Rate of stereo background events above a given mean size for data (the solid line) and MCs (dashed). 
The contribution of protons in the rate is shown with the dotted line, while helium and higher elements are shown with the dot-dashed line. 
}\label{fig_rate}
\end{figure}

We compare the rate of the reconstructed stereo background events above a given mean size (i.e. $(Size_1+Size_2)/2$) for the data and the MC samples (see Fig.~\ref{fig_rate}). 
We normalize the rate of the protons to the BESS measurement \citep{ya07}. 
At the energy close to the MAGIC threshold, the rate of Helium nuclei in the cosmic rays (CR) is comparable ($\sim 50-60\%$) to the rate of protons \citep{ho03}. 
Helium nuclei can be roughly treated as 4 nearly independent protons of 4 times smaller energy.
Thus the energy threshold for helium is higher than for protons and the rate of the MAGIC telescopes is dominated by the protons. 
Using proton and Helium MCs, and the CR composition \citep{ho03} we estimate, that the effect of elements  heavier than Helium on the rate of the background events registered by the MAGIC telescopes can be roughly approximated by considering an additional 60\% of Helium nuclei. 
After including all the components, the MC rate at low sizes agrees up to 20\% with the data. 
Differences at higher sizes are caused mostly by the small systematic errors in the spectral slope reconstruction of both MAGIC and BESS and the limited maximum energy of the proton and the Helium MCs. 

\begin{figure*}[t]
\centering 
\includegraphics[width=0.9\textwidth]{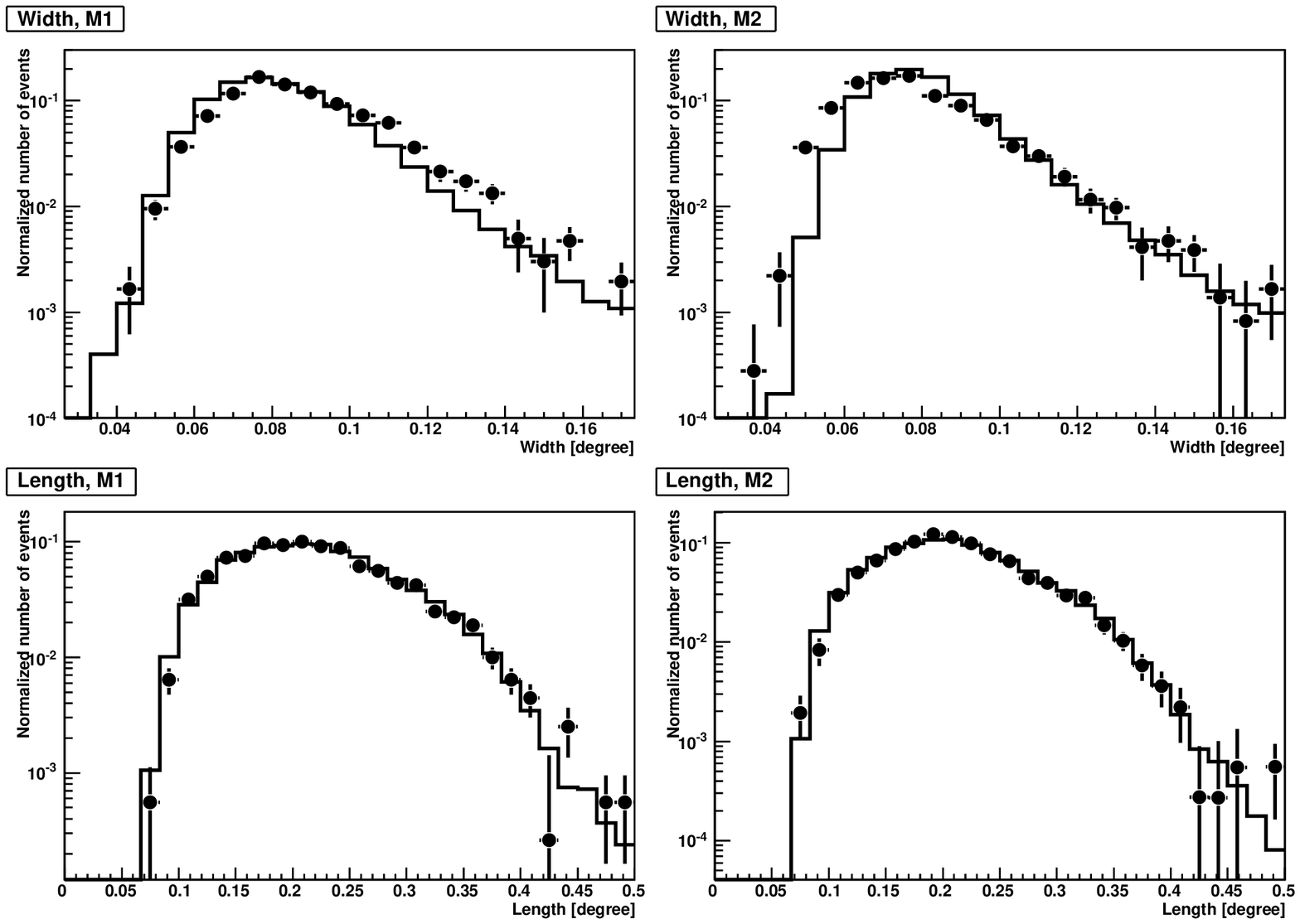}
\caption{
Comparison of the $Width$ (top panels) and the $Length$ (bottom) parameters for MAGIC I (left panels) and MAGIC II (right).
The solid line is computed with the MC gamma-rays, black points correspond to the gamma-rays excess, extracted with very loose cuts ($Hadronness<0.9$, $\theta^2<0.03$) from the data sample.  
Only events with $Size > 200\,$phe are used.  
}\label{fig_hillas}
\end{figure*}

The large data sample we use, and very good background reduction allows us to see a significant signal at medium energies even with very loose cuts, which corresponds to  a gamma-ray sample not biased by analysis cuts.
Therefore, we are able to extract distributions of individual parameters of gamma-rays from the data and compare them with the MC predictions. 
The comparison of the Hillas parameters $Width$ and $Length$ of the gamma-rays excess computed from the data sample with those from MCs is shown in Fig.~\ref{fig_hillas} for both telescopes separately.
Those parameters mostly agree between the MC gamma-rays and the gamma-ray excess extracted from the data.

\subsection{Energy resolution}
\begin{figure}[t]
\centering 
\includegraphics[width=0.49\textwidth]{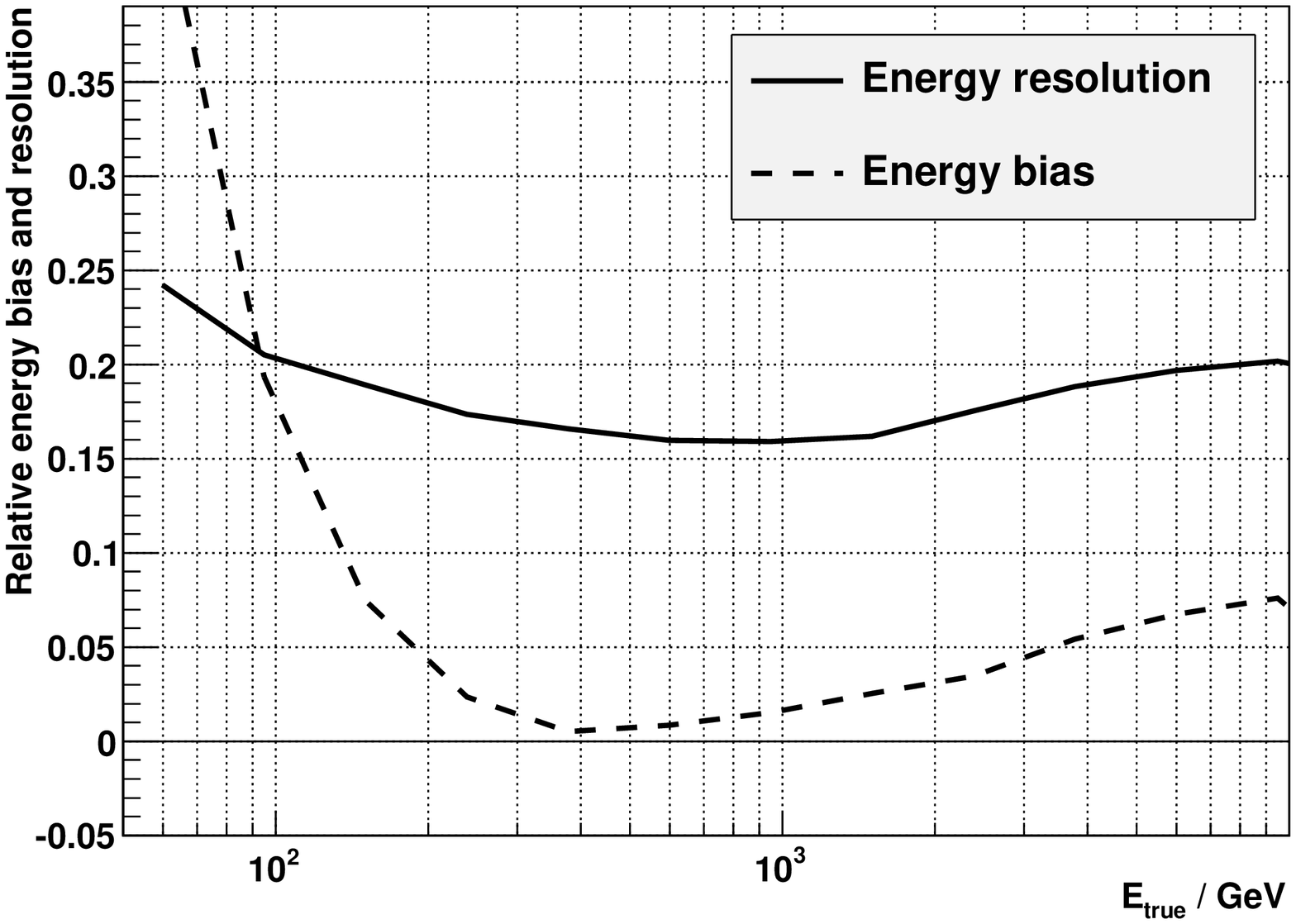}
\caption{
Energy resolution (solid line) and the bias (dashed) obtained with the MC simulations of gamma-rays.
Mild cuts are applied: $Hadronness<0.6$, $\theta^2<0.03$.
}\label{fig_erec}
\end{figure}

We evaluate the performance of the energy reconstruction with the help of gamma-ray MCs. 
We fill histograms of $(E_{rec}-E_{true})/E_{true}$ and fit them with a Gaussian distribution. 
The energy resolution can be calculated as the standard deviation of this distribution.
In addition, the bias introduced by the method is estimated as the mean value of the distribution.
The dependence of the energy resolution and the bias on the true energy of the gamma-rays is shown in Fig.~\ref{fig_erec}.
The energy resolution and bias weakly depends on the $Hadronness$ and $\theta^2$ cuts, usually improving for stronger cuts.

In the medium energy range (few hundred GeV) the energy resolution is as good as 16\%. 
For higher energies it is slightly worse due to the large fraction of truncated images, and showers with high $Impact$ parameters and worse statistics in the training sample.  
At low energies the energy resolution is worse, due to a lower photon number, higher relative noise, and worse estimation of the arrival direction, which spoils the precision of the $Impact$ parameter reconstruction.
In the medium energies the bias in the energy estimation is below a few per cent. 
At low energies ($\lesssim100\,$GeV) it is large due to the threshold effect. 
The bias is corrected in the analysis of the spectra by means of an unfolding procedure \citep{magic_unfolding}. 

\subsection{Angular resolution}
\begin{figure}
\centering 
\includegraphics[width=0.235\textwidth]{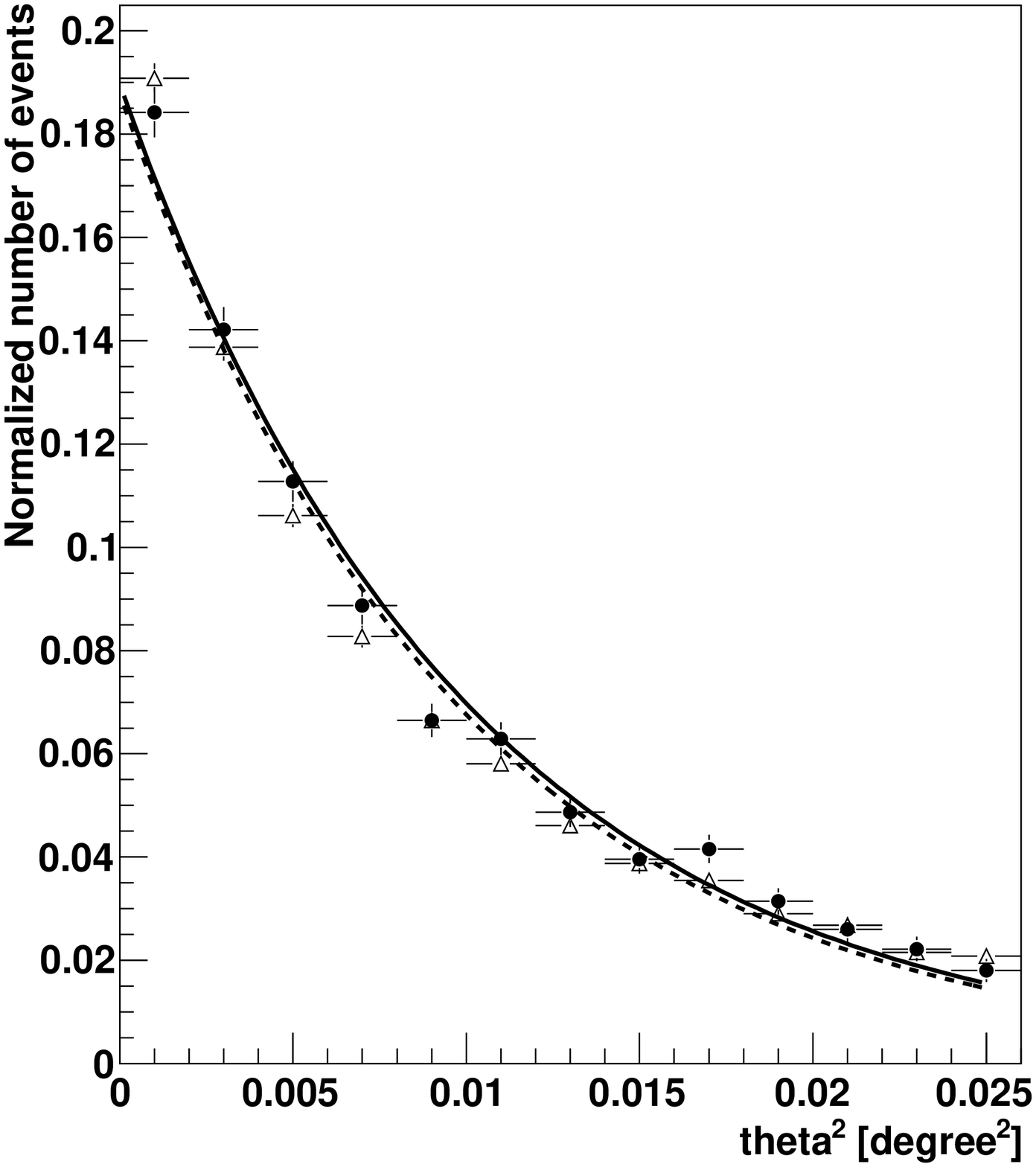}
\includegraphics[width=0.235\textwidth]{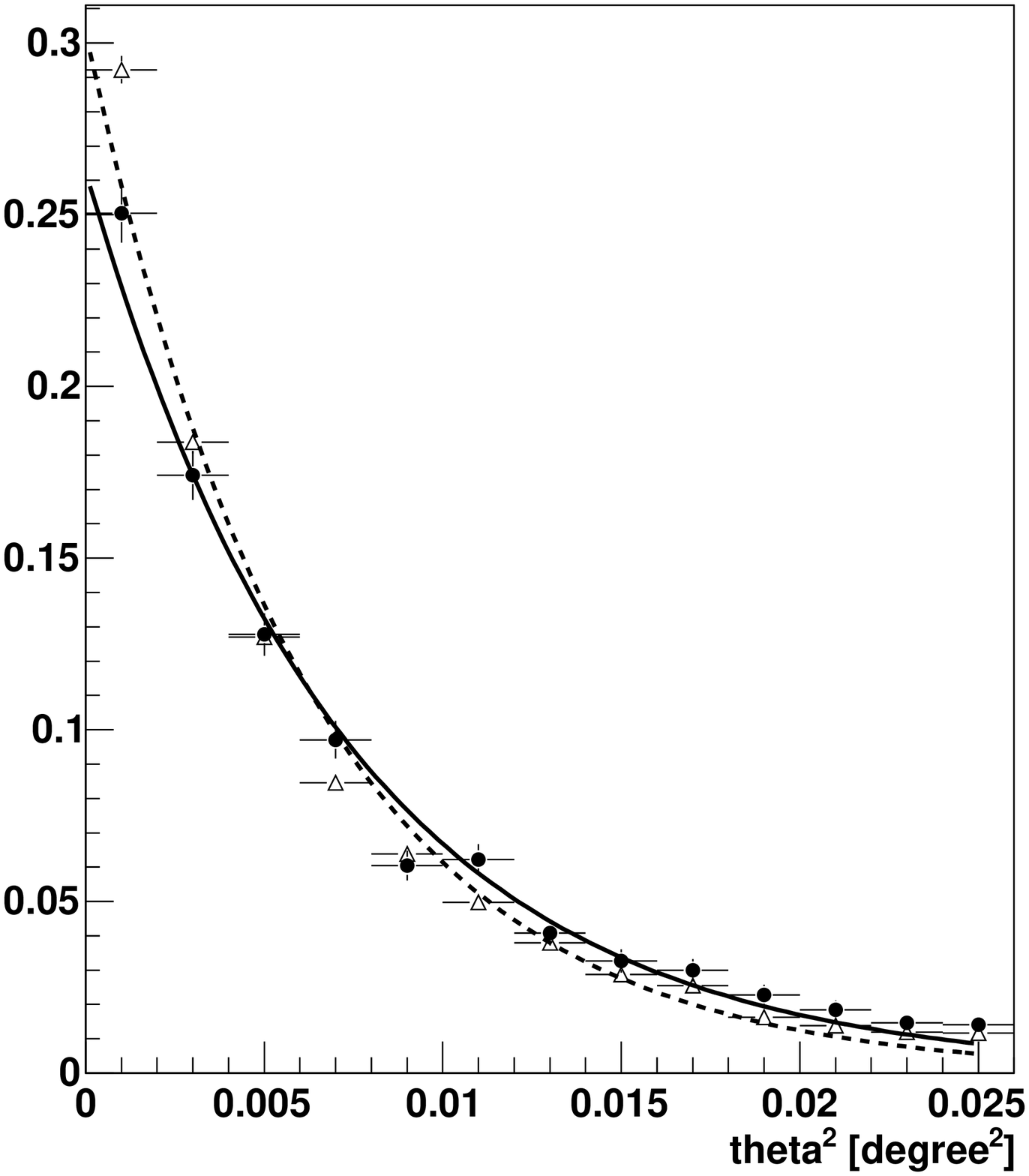}
\caption{
Theta$^{2}$ distribution of the excess above 100$\,$GeV (left panel) and above 300$\,$GeV (right), for the Crab Nebula data sample (solid circles) and MC gamma simulations (open triangles). 
Fits with a 2D Gaussian distribution are shown with a solid (for data) or dashed (for MCs) line. 
A small 2D Gaussian mispointing with $\sigma_{mis}=0.014^\circ$ was added to the simulations. 
}\label{fig_th2fit}
\end{figure}

We define the angular resolution as the standard deviation of the 2-dimensional Gaussian fitted to the distribution of the reconstructed event direction of the gamma-ray excess. 
This corresponds to a radius containing 39\% of the gamma-rays of a point like source. 
For completeness of the study we also calculate the 68\% containment radius.
In Fig.~\ref{fig_th2fit} we show the $\theta^{2}$ distribution of the gamma-ray excess above two different energy thresholds for data and MC simulations. 
The distributions for $\theta^2<0.025$ can be reasonably well fitted with a single Gaussian.  
The distribution in a broader signal range is better described by a double Gaussian.
In the lower energies MC and data are consistent. 
For higher energies, due to small imperfections not included in the MC simulations, the MC distribution is narrower resulting in $\lesssim20\%$ higher number of events in the first bin ($\theta^2<0.002$). 
This does not influence the normal analysis, since the usual cut in $\theta^2$ is a factor of 5--10 larger, and the signal is recovered. 

\begin{figure}
\centering 
\includegraphics[width=0.49\textwidth]{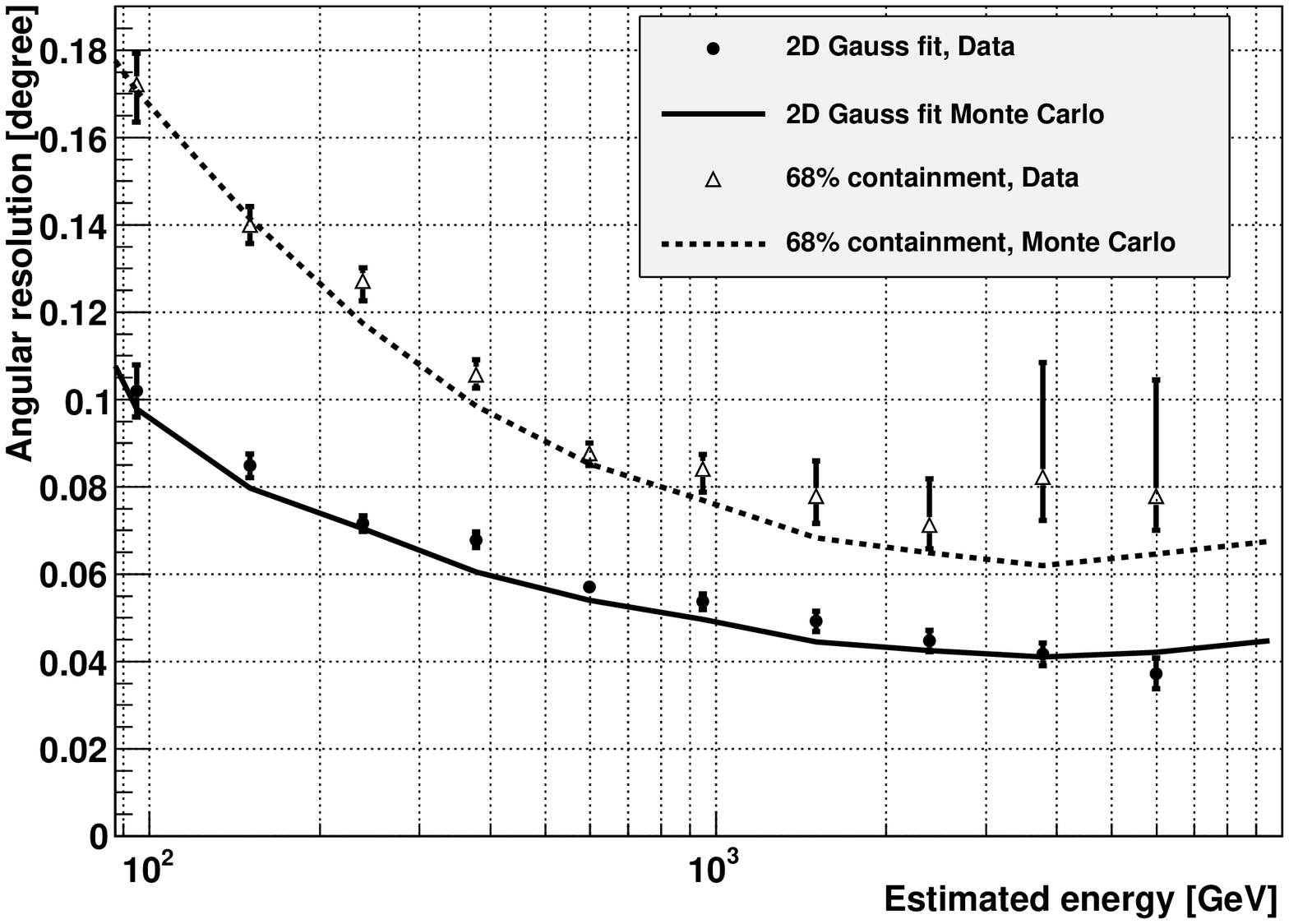}
\caption{
Angular resolution as a function of the estimated energy obtained with the Crab Nebula data sample (points) and compared with the MC simulations (lines) for the 2D Gaussian fit (solid line, full circles) and the 68\% containment radius (dashed line, empty triangles).
A small 2D Gaussian mispointing with $\sigma_{mis}=0.014^\circ$ was added to the simulations. 
}\label{fig_angres}
\end{figure}

The angular resolution obtained from MC and data are consistent (see Fig.~\ref{fig_angres}), with just a small discrepancy mostly visible at the highest energies ($\gtrsim 1\,$TeV). 
The stereo DISP RF method allows us to obtain a very good angular resolution of $\sim0.07^\circ$ at 300$\,$GeV, and even better at higher energies.

\subsection{Spectrum of the Crab Nebula}

\begin{figure}
\centering 
\includegraphics[width=0.49\textwidth]{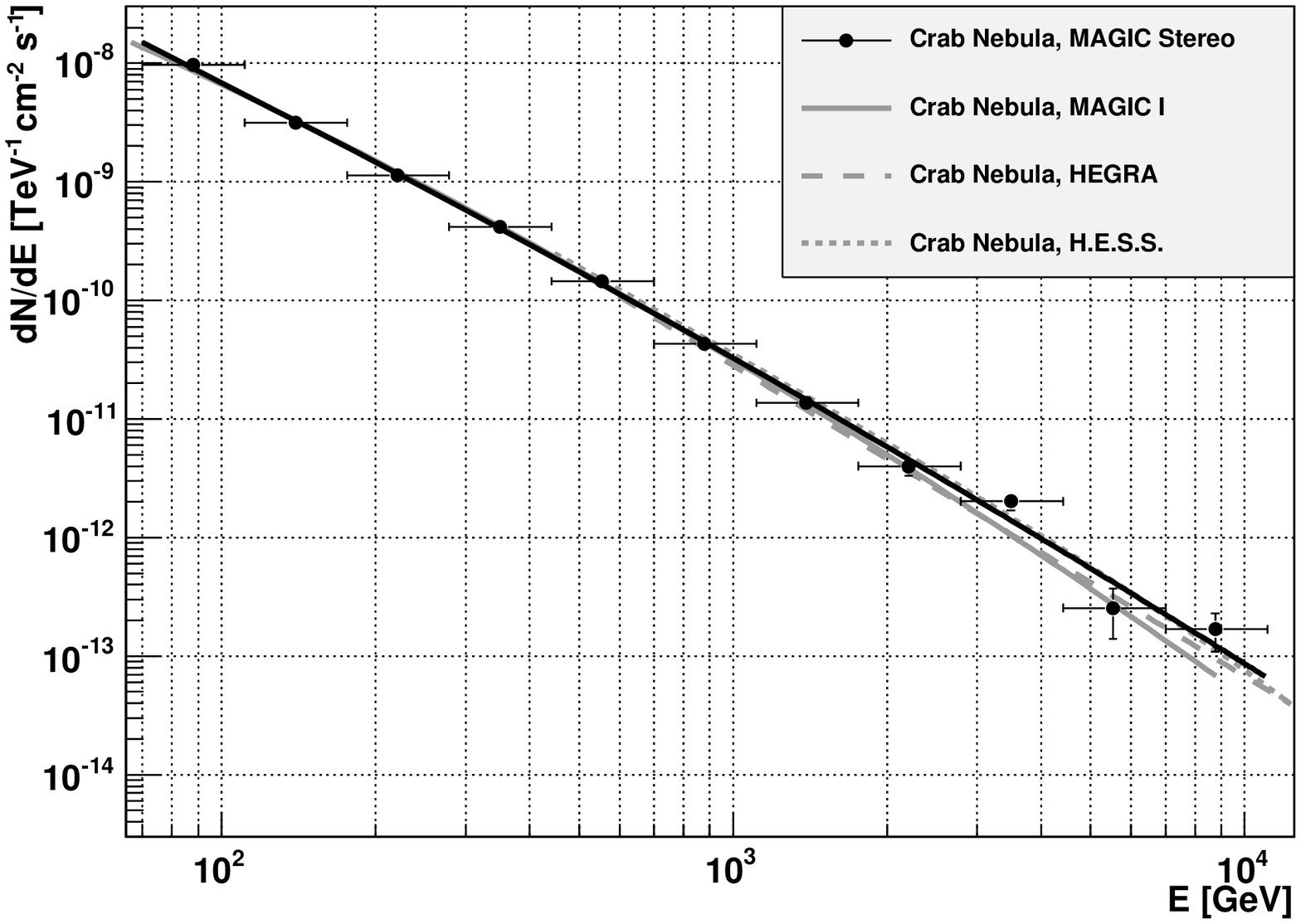}\\
\includegraphics[width=0.49\textwidth]{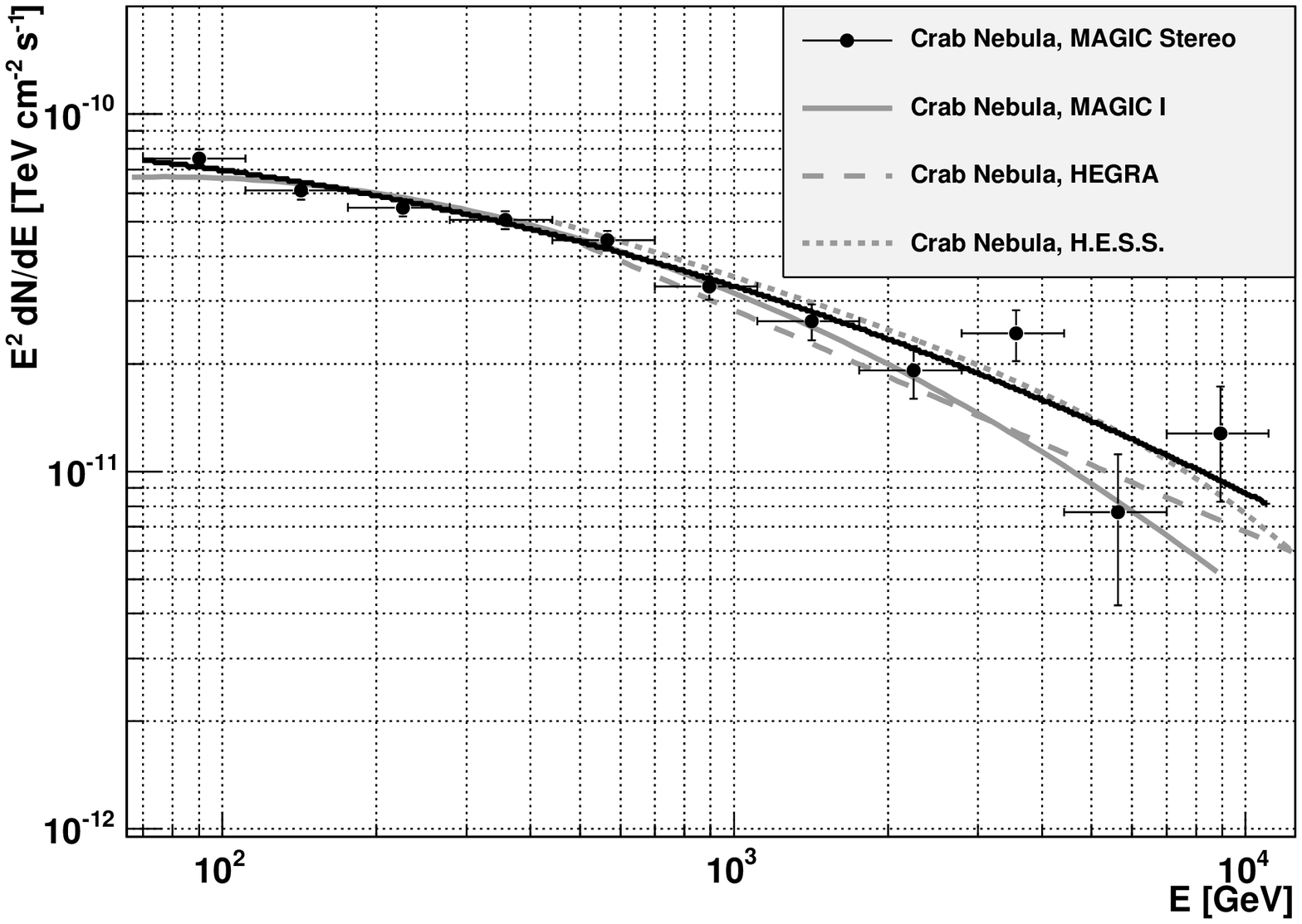}
\caption{
Spectrum (top panel) and the SED (bottom) of the Crab Nebula obtained with the MAGIC Stereo system (black) compared with other experiments (gray curves): MAGIC~I (solid, \citealp{magic_crab}), HEGRA (dashed, \citealp{ah04}), and H.E.S.S. (dotted, \citealp{ah06}). 
The vertical error bars show statistical errors. 
}\label{fig_spectrum}
\end{figure}

\komm{In order to minimize the systematic errors we apply $Hadronness$ and $\theta^2$ cuts with high gamma-ray efficiency for the reconstruction of spectrum of the Crab Nebula.}
The spectrum and the spectral energy distribution (SED) of the Crab Nebula obtained from the analyzed sample is presented in Fig.~\ref{fig_spectrum}.
The spectrum in the energy range 70$\,$GeV -- 11$\,$TeV can be fitted with a curved power-law:
\begin{equation}
\frac{dN}{dE}=f_0( E/ 300\ \mathrm{GeV}\,) ^{a+b\mathrm{\log}_{10}( E/ 300\,\mathrm{GeV}\,)}\ 
\mathrm{ [cm^{-2} s^{-1} TeV^{-1}]}, \label{eq_spectrum}
\end{equation} 
where the obtained parameters of the fit are:
$f_0 =( 5.8\pm 0.1_{\mathrm{stat}}) \times 10^{-10}$, 
$a=-2.32\pm 0.02_{\mathrm{stat}}$, 
and $b=-0.13\pm 0.04_{\mathrm{stat}}$.

The spectrum obtained with the MAGIC Stereo observations agrees within 20-30\% with the previous measurements of the Crab Nebula.
The curvature seems to be smaller than reported in \citet{magic_crab}, however the effective spectral slope is still within the systematic and statistical error.
Note moreover, that the fitting range is slightly different than in \citet{magic_crab}. 
The spectrum in a broader energy range, together with its scientific implications will be discussed in another paper (in preparation), which exploits a dedicated analysis for the lowest and the highest energies.

\subsection{Integral sensitivity}
In order to allow for a fast reference and comparison with other experiments we calculate the sensitivity of the MAGIC telescopes according to a number of commonly used definitions. 
For a weak source and perfectly known background, the significance of a detection can be calculated with a simplified formula $N_{\rm excess}/\sqrt{N_{\rm bgd}}$, where $N_{\rm excess}$ is the number of the excess events, and $N_{\rm bgd}$ is the estimation of the background. 
We compute the sensitivity $S_{Nex/\!\sqrt{Nbkg}}$ as the flux of a source giving $N_{\rm excess}/\sqrt{N_{\rm bgd}}=5$ after 50$\,$h of effective observation time.

In the calculation of the sensitivity, $S_{Nex/\!\sqrt{Nbkg},\, {\rm sys}}$, we are applying also two additional conditions $N_{\rm excess}>10$, $N_{\rm excess}> 0.05 N_{\rm bgd}$.
The second condition protects against small systematic discrepancies between the ON and the OFF, which may mimic a statistically significant signal if the residual background rate is large. 
Finally, the sensitivity can also be calculated using the \citet{lm83}, eq.~17 formula for significance, which is the standard method in the VHE gamma-ray astronomy for the calculation of the significance. 

\begin{figure}
\centering 
\includegraphics[width=0.49\textwidth]{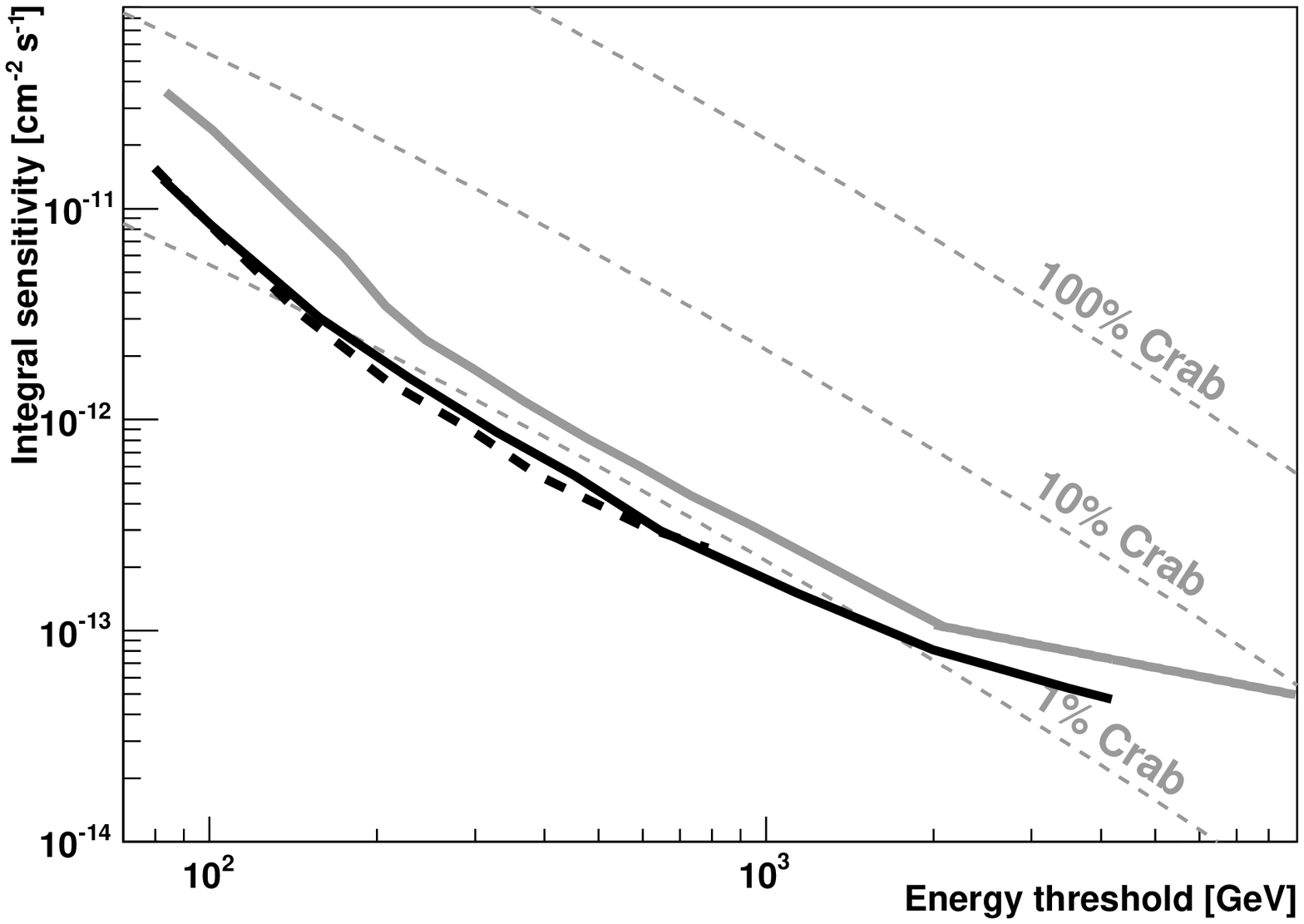}
\caption{
Integral sensitivity of the MAGIC Stereo system, i.e. the flux of a source above a given energy for which $N_{\rm excess}/\sqrt{N_{\rm bgd}}=5$ after $50\,\mathrm{h}$ of effective observation time for MAGIC Stereo (black, solid -- data, dashed -- MCs) compared with the MAGIC~I telescope (solid gray line).
For comparison, fractions of the integral Crab Nebula spectrum are plotted with the thin, dashed, gray lines. 
}\label{fig_sens}
\end{figure}

The integral sensitivity of the MAGIC telescopes working in stereo mode is shown in Fig.~\ref{fig_sens} and the values both in Crab Units (C.U.) and in absolute units are summarized in Table~\ref{tab_sens}.
\begin{table*}[t]
\centering
\begin{tabular*}{0.7\textwidth}{@{\extracolsep{\fill}} c|c|c|c|c|c} 
$E_{\rm thresh.}$ & $S_{Nex/\!\sqrt{Nbkg}}$ & $S_{Nex/\!\sqrt{Nbkg},\, {\rm sys}}$ & $S_{\rm Li\&Ma, 1Off}$ & $S_{\rm Li\&Ma, 3Off}$ & $S_{Nex/\!\sqrt{Nbkg}}$  \\
~[GeV] & [\%C.U.] & [\%C.U.] & [\%C.U.] & [\%C.U.]&[$10^{-13}$ $\rm cm^{-2} s^{-1}$] \\\hline\hline
82.5 & $1.99 \pm 0.03$ & $4.7 \pm 0.04$ & $2.84 \pm 0.019$ & $2.314 \pm 0.016$ & $138 \pm 2$ \\ \hline
100 & $1.56 \pm 0.02$ & $2.41 \pm 0.03$ & $2.24 \pm 0.02$ & $1.818 \pm 0.018$ & $84.1 \pm 1.3$ \\ \hline
158 & $1.02 \pm 0.02$ & $1.02 \pm 0.02$ & $1.48 \pm 0.03$ & $1.2 \pm 0.02$ & $30.4 \pm 0.6$ \\ \hline
229 & $0.87 \pm 0.02$ & $0.87 \pm 0.02$ & $1.29 \pm 0.03$ & $1.03 \pm 0.02$ & $15.7 \pm 0.4$ \\ \hline
328 & $0.79 \pm 0.03$ & $0.79 \pm 0.03$ & $1.23 \pm 0.05$ & $0.97 \pm 0.04$ & $8.7 \pm 0.4$ \\ \hline
452 & $0.78 \pm 0.04$ & $0.78 \pm 0.04$ & $1.25 \pm 0.06$ & $0.98 \pm 0.05$ & $5.4 \pm 0.3$ \\ \hline
646 & $0.72 \pm 0.06$ & $0.72 \pm 0.06$ & $1.23 \pm 0.08$ & $0.95 \pm 0.07$ & $3 \pm 0.3$ \\ \hline
1130 & $0.86 \pm 0.06$ & $0.86 \pm 0.06$ & $1.67 \pm 0.07$ & $1.23 \pm 0.06$ & $1.51 \pm 0.11$ \\ \hline
2000 & $1.12 \pm 0.14$ & $1.12 \pm 0.14$ & $2.67 \pm 0.16$ & $1.85 \pm 0.13$ & $0.81 \pm 0.1$ \\ \hline
2730 & $1.5 \pm 0.3$ & $1.58 \pm 0.15$ & $4.3 \pm 0.3$ & $2.8 \pm 0.3$ & $0.64 \pm 0.12$ \\ \hline
3490 & $1.8 \pm 0.4$ & $2.3 \pm 0.3$ & $5.9 \pm 0.5$ & $3.8 \pm 0.4$ & $0.53 \pm 0.11$ \\ \hline
4180 & $2.3 \pm 0.5$ & $2.9 \pm 0.4$ & $7.5 \pm 0.6$ & $4.8 \pm 0.5$ & $0.47 \pm 0.1$ \\ \hline

\end{tabular*}
\caption{
Integral sensitivity obtained with the Crab Nebula data sample above the energy threshold $E_{\rm thresh.}$.
The sensitivity is calculated as $N_{\rm excess}/\sqrt{N_{\rm bgd}}=5$ after 50$\,$h of effective time ($S_{Nex/\!\sqrt{Nbkg}}$), or with additional conditions $N_{\rm excess}>10$, $N_{\rm excess}> 0.05 N_{\rm bgd}$ ($S_{Nex/\!\sqrt{Nbkg},\, {\rm sys}}$).
The sensitivity calculated to obtain 5$\sigma$ significance after 50$\,$h of effective time according to \citet{lm83} with an assumption of 1 or 3 background regions is shown in $S_{\rm Li\&Ma, 1Off}$ and $S_{\rm Li\&Ma, 3Off}$ columns respectively. 
}\label{tab_sens}
\end{table*}
Above a few hundred GeV the MAGIC Stereo sensitivity is a factor of 2 better than the one of the MAGIC~I telescope.
The improvement is even larger (by a factor of about 3) at lower energies. 
Note that two identical telescopes working completely independently would have a sensitivity only a factor $\sqrt{2}$ better compared to a single one. 
The improvement by a factor of 2-3 therefore comes from the efficient usage of the stereo parameters in the analysis and the intrinsic reduction of the muon background in stereo systems. 

In Fig.~\ref{fig_sens} we also show the sensitivity obtained with sets of gamma-ray, proton, helium and electron MC simulations. 
We scale up the proton MCs by 20\% to take into account the difference in the rate between the data and the MCs (see Section~\ref{sec:parameters}). 
The flux of the electrons is much smaller than the one of the protons, and the spectrum is steeper \citep{ah09}.
Nevertheless, due to their similarity to the gamma-ray showers, they can constitute a significant part of the background after gamma/hadron separation at medium and high energies (about 15\% at 200$\,$GeV).
The sensitivity obtained from the Crab data set is in agreement with the MC predictions.

\subsection{Differential sensitivity}\label{sec_diffsens}

\begin{figure}
\centering 
\includegraphics[width=0.49\textwidth]{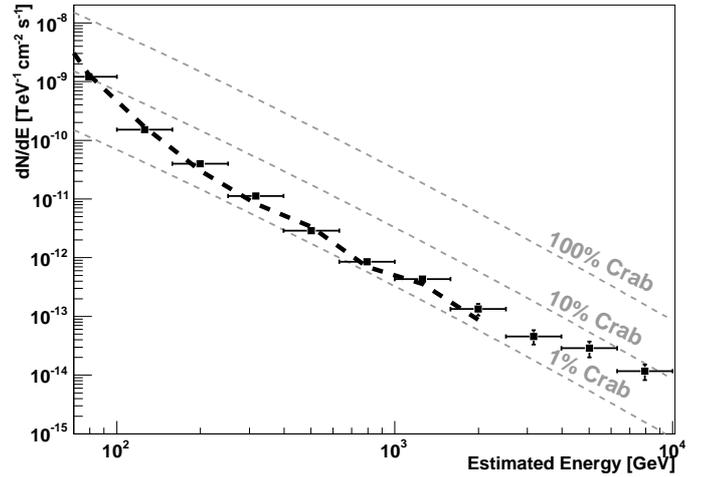}
\caption{
Differential sensitivity of the MAGIC Stereo system.
We compute the flux of the source in a given energy range for which $N_{\rm excess}/\sqrt{N_{\rm bgd}}=5$, $N_{\rm excess}>10$, $N_{\rm excess}> 0.05 N_{\rm bgd}$ after 50$\,$h of effective time for the data (black points), and the MC (dashed line).
The horizontal error bars represent the size of the bins in estimated energy (5 bins per decade). 
For comparison, fractions of the Crab Nebula spectrum are plotted with the thin dashed gray lines. 
}\label{fig_diffsens}
\end{figure}

To evaluate the performance of the MAGIC telescopes for sources with a substantially different spectral shape compared to the Crab Nebula, we compute also the differential sensitivity, i.e. we investigate the sensitivity in narrow bins of energy (5 bins per decade).
In order to derive optimal cuts in $Hadronness$ and $\theta^2$ in each energy bin, we divide the sample into two roughly equal subsamples. 
In each energy bin we perform a scan of cuts on the \komm{training} subsample, and apply the best cuts to the second sample obtaining the sensitivity (see Fig.~\ref{fig_diffsens} and Table~\ref{tab_diffsens}).

\begin{table}[t]
\centering
\small
\begin{tabular}{c|c|c|c}
$E_{\rm min}$ & $E_{\rm max}$ & $S_{Nex/\!\sqrt{Nbkg},\, {\rm sys}}$ & $S_{Nex/\!\sqrt{Nbkg},\, {\rm sys}}$ \\
~[GeV] & ~[GeV] & [\%C.U.] &[$10^{-12}$ $\rm TeV^{-1} cm^{-2} s^{-1}$]\\\hline\hline
63.1 & 100 & $10.5 \pm 1.3$ & $1210 \pm 150$ \\ \hline
100 & 158 & $3.65 \pm 0.23$ & $152 \pm 9$ \\ \hline
158 & 251 & $2.7 \pm 0.19$ & $40 \pm 3$ \\ \hline
251 & 398 & $2.2 \pm 0.25$ & $11.3 \pm 1.3$ \\ \hline
398 & 631 & $1.67 \pm 0.22$ & $2.9 \pm 0.4$ \\ \hline
631 & 1000 & $1.49 \pm 0.2$ & $0.85 \pm 0.11$ \\ \hline
1000 & 1580 & $2.3 \pm 0.4$ & $0.43 \pm 0.07$ \\ \hline
1580 & 2510 & $2.3 \pm 0.5$ & $0.13 \pm 0.03$ \\ \hline
2510 & 3980 & $2.5 \pm 0.7$ & $0.046 \pm 0.013$ \\ \hline
3980 & 6310 & $5.3 \pm 1.6$ & $0.029 \pm 0.009$ \\ \hline
6310 & 10000 & $7.4 \pm 2.2$ & $0.012 \pm 0.003$ \\ \hline
\end{tabular}
\caption{
Differential sensitivity obtained with the Crab Nebula data sample.
The sensitivity in each energy bin $(E_{\rm min}, E_{\rm max})$ is calculated as $N_{\rm excess}/\sqrt{N_{\rm bgd}}=5$, $N_{\rm excess}>10$, $N_{\rm excess}> 0.05 N_{\rm bgd}$ after 50$\,$h.
}\label{tab_diffsens}
\end{table}

At medium energies, the sensitivity is very good (about 1.5-2.5\% C.U.).
Even below 100$\,$GeV, in the energy regime exclusive to the MAGIC telescopes among all the current IACTs, we obtain a good sensitivity of $\sim$ 10\% C.U.
As in the case of the integral sensitivity, the differential sensitivity derived with the help of MC simulations is in agreement with the one obtained from the data sample.

\subsection{Off-axis performance}
Most of the observations of the MAGIC telescopes are performed in the wobble mode with an offset of $0.4^\circ$. 
However, serendipitous sources can occur in the FoV of MAGIC at an arbitrary angular offset from the viewing direction (see e.g. detection of IC~310, \citealp{magic_ic310}).
Therefore, we also study \komm{the angular resolution and} the sensitivity of MAGIC at different offsets from the center of the FoV. 
Dedicated observations of the Crab Nebula were performed at a wobble offset of \komm{$\xi=$}$0.2^\circ$, $0.3^\circ$, $0.6^\circ$, $1^\circ$, and $1.4^\circ$.

\komm{
Since the presented here analysis is optimized for sources observed at the offset of $0.4^\circ$ from the camera centre, the best angular resolution is obtained for this angle (see Fig.~\ref{fig_offset_res}).
\begin{figure}[t]
\centering 
\includegraphics[width=0.49\textwidth]{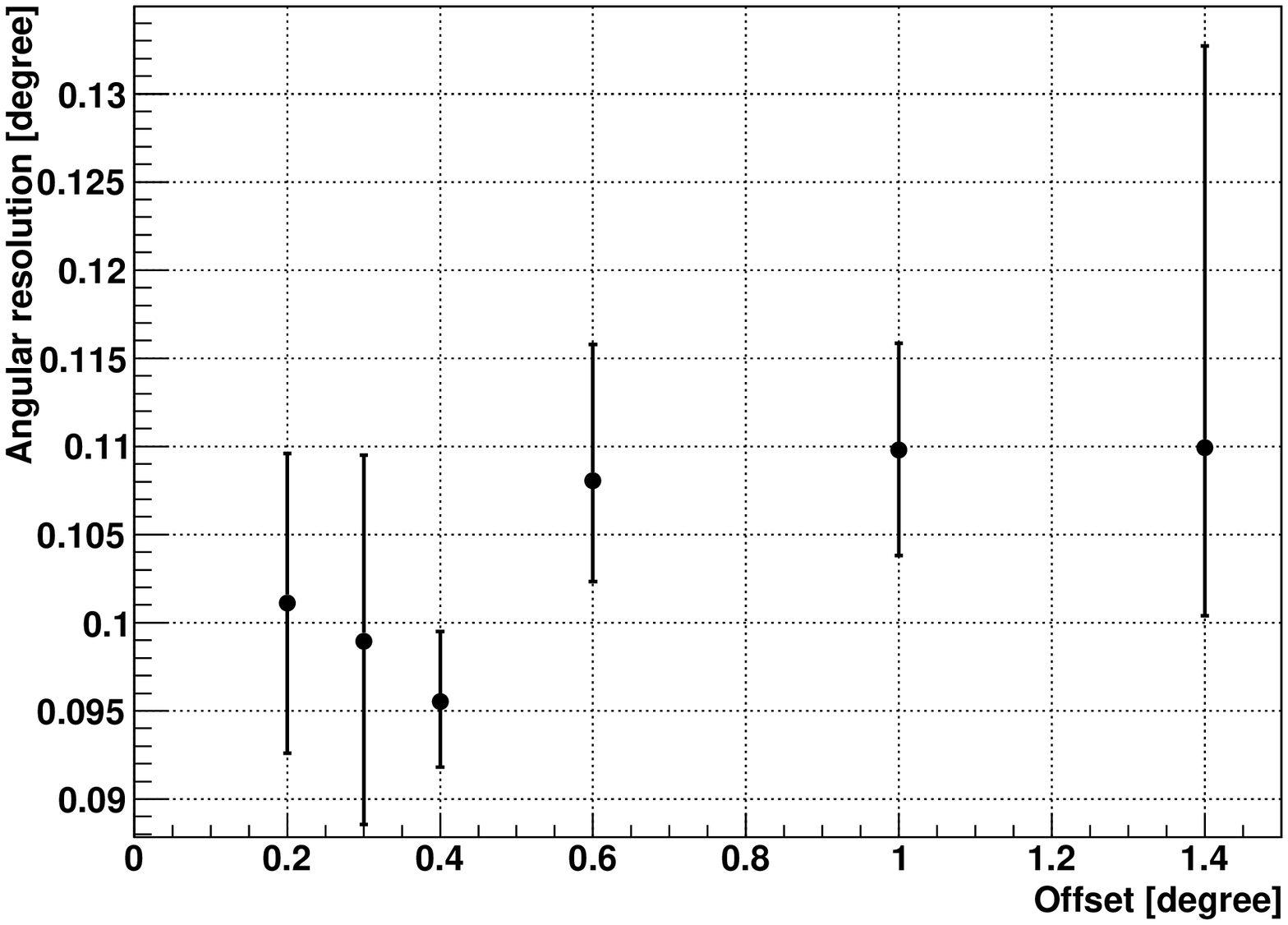}
\caption{
\komm{
Angular resolution (defined as the 68\% containment radius) for gamma-rays with energy above 200 GeV for the observation of a source at a different offset from the center of the camera.
Each point is obtained with a small, dedicated sample of Crab Nebula data.
}
}\label{fig_offset_res}
\end{figure}
For larger offsets the angular resolution is slightly spoiled ($\sim 15\%$ at $\xi = 1^\circ$), possibly due to higher influence of the optical aberrations. 
MC simulations show a small (5\% of the offset angle) radial bias in the estimated source position. 
Since normally at least two different pointing positions are used for each source, the impact of this bias on the data is reduced. 
}

\begin{figure}[t]
\centering 
\includegraphics[width=0.49\textwidth]{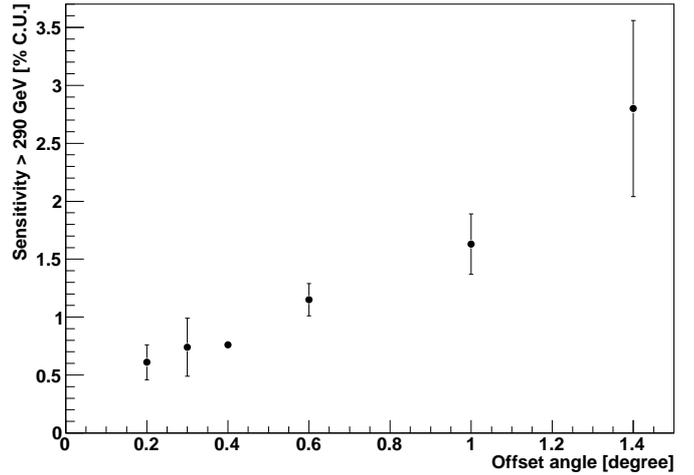}
\caption{
Integral sensitivity above 290$\,$GeV for the observation of a source at a different offset from the center of the camera. 
Each point is obtained with a small, dedicated sample of Crab Nebula data.
}\label{fig_offset}
\end{figure}

The sensitivity above 290$\,$GeV for different offsets is plotted in Fig.~\ref{fig_offset}. 
The values of the sensitivity and the effective observation times of the data samples taken at different offsets are summarized in Table~\ref{tab_offset}.
\begin{table}[t]
\centering
\small
\begin{tabular}{c|c|c}
$\xi [^\circ]$ & time[h] & $S_{Nex/\!\sqrt{Nbkg}}$ [\% C.U.] \\\hline\hline
0.2 & 0.46 & 0.61 $\pm$ 0.15 \\ \hline
0.3 & 0.23 & 0.74 $\pm$ 0.25 \\ \hline
0.4 & 8.7 & 0.76 $\pm$ 0.03 \\ \hline
0.6 & 0.89 & 1.15 $\pm$ 0.14 \\ \hline
1 & 1.55 & 1.63 $\pm$ 0.26 \\ \hline
1.4 & 0.84 & 2.8 $\pm$ 0.76 \\ \hline
\end{tabular}
\caption{
Integral sensitivity $S_{Nex/\!\sqrt{Nbkg}}$ above 290$\,$GeV (calculated as $N_{\rm excess}/\sqrt{N_{\rm bgd}}=5$ after 50$\,$h) obtained with the Crab Nebula data samples for different offsets $\xi$ of the source from the center of the camera.
}\label{tab_offset}
\end{table}
The sensitivity of the MAGIC telescopes is quite good if the projected position of the source is within the inner part of the cameras.
The sensitivity degrades significantly for sources located at a larger distance from the camera center (about a factor of 2 for $1^\circ$ offset).

\section{Study of the systematic uncertainties}\label{sec:systematics}
Due to the complexity of the IACT technique, there are many factors which are only known with some uncertainties, thus contributing to systematic errors (see Table~\ref{tab_syst}).
We consider the systematic errors on the gamma-ray collection efficiency (i.e. on the absolute flux level), on the absolute light scale, and on the reconstructed spectral slope.

\begin{table}[t]
\centering
\small
\begin{tabular}{l|r}
Systematic effect & Uncertainty  \\\hline\hline
F-Factor & 10\% ES \\ \hline
atmospheric transmission & $\lesssim$10\% ES \\ \hline
mirror reflectivity & 8\% ES \\ \hline
PMT electron collection efficiency & 5\% ES \\\hline
light collection in a Winston Cone & 5\% ES \\\hline
PMT quantum efficiency & 4\% ES\\\hline
signal extraction & 3\% ES \\ \hline
temperature dependence of gains & 2\% ES \\ \hline
charge flat-fielding & 2-8\% ES FN \\ \hline
analysis and MC discrepancies & $\lesssim$10-15\% FN \\ \hline
background subtraction & 1-8\% FN \\ \hline
broken channels/pixels & 3\% FN \\ \hline
mispointing & 1-4\% FN \\ \hline
NSB & 1-4\% FN \\ \hline
trigger & 1\% FN \\ \hline
unfolding of energy spectra & 0.1 SL \\ \hline
non-linearity of readout & 0.04 SL \\ \hline
\end{tabular}
\caption{ 
Estimated values of the main sources of systematic uncertainties. 
They can affect the energy scale (ES), flux normalization (FN) and the spectral slope (SL). 
See text for the detailed explanation.
}\label{tab_syst}
\end{table}

\subsection{Atmosphere}
The Earth's atmosphere has to be considered part of an IACT detector.
Changing atmospheric conditions, small deviations of the density profile from the one assumed in simulations, as well as non-perfectly known atmospheric transmission due to Mie scattering introduce additional uncertainties in the energy scale and a small effect on the spectral slope reconstruction.
We estimated the systematic error in the energy scale due to atmospheric parameters to be $\lesssim 10\%$. 

\subsection{Mirrors and night sky background}
An uncertainty in the amount of light focused by the mirrors can be caused by non-perfect knowledge of the reflectivity of the mirrors, in particular short-term variations of the dust deposit.  
\komm{In addition to this}, due to malfunctions of the AMC system, the total useful mirror area can vary from one night to another. 
Using the measurements of the reflected star images and the analysis of the muon images, we estimated that those effects produce a systematic error on the energy scale of about $\sim$ 8\%. 

The level of the night sky background (NSB) can vary from one observation to another. 
In particular, galactic sources usually have a higher NSB than extragalactic ones.
Also, observations in twilight and moonlight conditions exhibit higher NSB values. 
High NSB increases the fluctuations of the signal in a pixel. 
Thus it can spoil the precision of the estimation of the Hillas parameters, and hence lower the acceptance for gamma-rays. 
%%We investigated this effect generating dedicated MCs with a 30\% higher NSB than a typical galactic region of sky and found that the effect on the gamma-ray collection efficiency is up to 4\% at low energies ($\sim 100\,$GeV) and negligible ($\lesssim1\%$) above 300$\,$GeV.
We investigated this effect generating dedicated MCs with a 30\% higher NSB than a typical galactic region of sky.
\komm{Such an increase covers typical changes of the mean NSB observed in different regions of the sky in dark conditions, and a mild moonlight (moon phase $\lesssim 20\%$).}
We found that the effect on the gamma-ray collection efficiency is up to 4\% at low energies ($\sim 100\,$GeV) and negligible ($\lesssim1\%$) above 300$\,$GeV.

\subsection{Cameras and readout}
An important systematic error in the absolute energy scale comes from uncertainties of the conversion coefficient of photons to detectable photoelectrons.
It contains uncertainty in the light collection in the Winston Cone ($\sim 4\%$), electron collection efficiency of the first dynode ($\sim 5\%$), quantum efficiency of the PMT ($\sim 4\%$) and finally the F-Factor value of the PMTs ($\sim 10\%$).
While it is difficult to disentangle and correct the individual components of this energy scale uncertainty, by using an absolute muon calibration \citep{pu03}, and intertelescope cross-calibration we are able to obtain much more precise photon to phe conversion factors.  

In addition, a $\sim2\%$ effect in the energy scale can be attributed to the temperature dependence of the gains.
In order to make the response of the camera homogeneous, we perform the so-called flat-fielding.
Flat-fielding equalizes the product of the quantum efficiency of the pixel for the wavelength of the light pulser with the gain of the PMT-FADC chain. 
The part of the signal coming to the trigger branch is thus only partially flatfielded, moreover temperature drifts can influence the flat-fielding. 
Those effects, and the fact that the flat-fielding of the signal chain is done only at one wavelength, can produce a $\sim6-8\%$ systematic uncertainty in the energy scale and the event rates, mostly for the small events, and a smaller effect ($\lesssim2\%$) at higher energies. 
With over 1600 channels in total it is natural that a small number of them ($\lesssim 10$) at a given moment is unusable, due to e.g. a malfunctioning PMT or readout electronics. 
Since we interpolate the signal in those channels in the analysis, the systematic effect on the energy scale is negligible. 
However the interpolation procedure can lead to a loss of about 3\% of gamma-ray efficiency. 
Non-linearities in the analog signal chain and the small residual non-linearity of the DRS2 chip can produce a systematic uncertainty of about 0.04 in the spectral index. 

\subsection{Trigger}
The readout of MAGIC~II introduces a dead time of about $10\%$ for typical trigger rate of events. 
The dead time is corrected in the analysis during the calculation of the effective observation time by means of an exponential fit to the distribution of the time differences to the previous triggered event. 
The remaining error is negligible ($\lesssim 1$\%).
Also, the inefficiencies of the trigger systems of the MAGIC telescopes are negligible.

\subsection{Signal extraction}
We investigate the systematic uncertainties induced by the signal extraction in MAGIC~II by varying the size of the extraction window. 
We find that the difference in the reconstructed number of photoelectrons is $\lesssim 3\%$. 
It is similar to the corresponding value of MAGIC~I.

\subsection{Mispointing}
Mispointing of one or both telescopes can influence the analysis. 
Not only the $\theta^2$ distribution becomes broader, but the relative pointing differences between MAGIC~I and MAGIC~II spoil the reconstruction of the stereo parameters. 
The typical mispointing of the individual MAGIC telescopes is $\lesssim0.02^\circ$ \citep{magic_drive, magic_halo}.
The final, post-analysis mispointing is slightly higher $\lesssim0.025^\circ$ since it includes reconstruction biases. 
We performed dedicated MC simulations and conclude that the systematic error on the gamma-ray efficiency due to mispointing is $\lesssim4\%$

\subsection{Background subtraction}
Due to dead pixels and stars in the field of view of the telescopes, the distribution of the events on the camera can become partially inhomogeneous (i.e. it loses its rotational symmetry).
On top of this, the stereo trigger produced a natural inhomogeneity. 
As a first approximation, the stereo FoV can be treated as a non-circular \komm{overlapping} region of two cones (size of each determined by the size of the camera trigger regions) originating from both telescopes. 
This results in an enhanced amount of events in the direction determined by the position of the second telescope and can lead to a 10-15\% variation in the background estimation. 
Such an error would significantly influence the reconstructed flux and spectral index of a weak source.
However, since the MAGIC data are usually taken in wobble mode, alternating the source and anti-source position, the systematic uncertainty of the background estimation is strongly reduced down to $\lesssim 2\%$. 
This procedure naturally does not work for serendipitous discoveries. 
Nevertheless, even in this case a comparable decrease of the systematic error is obtained via renormalization of the source and anti-source $\theta^2$ plots in the off-source region or taking the background estimate from the geometrically equivalent position in the FoV of the other wobble position. 
The effect of inhomogeneity is mostly pronounced at the lowest energies, where usually the excess is smaller than the background.
Thus, as an example, a weak source with an excess being just 25\% of the background, the induced error of the flux estimation can be up to $\sim 8$\%.

\subsection{Analysis}
Because of small remaining discrepancies between data and MCs the reconstructed spectra depend slightly ($\lesssim 10\%$ at medium energies, and $\lesssim 15\%$ at low energies) on the value of cuts used for extracting the gamma-ray signal.  

Also, the unfolding procedure, which is needed to correct for the finite energy resolution and the energy bias at the lowest energies, introduces additional systematic uncertainties.
We can estimate them by comparing results obtained with different regularization methods \citep{magic_unfolding}.
We obtain that for a typical power-law spectrum the spread of photon indices is 0.1.

\subsection{Total systematic uncertainty} 
As some of the systematic errors depend strongly on the energy of the gamma-ray showers, we summarize separately the systematic errors for low ($\lesssim$ 100$\,$GeV) and for medium energies ($\gtrsim300\,$GeV).
As the individual errors are mostly independent from each other we add them in quadrature (as in \citealp{magic_crab}).
The absolute muon calibration and intertelescope cross-calibration allows us to remove most of the errors connected with the photon to phe conversion.
Thus, the energy scale of the MAGIC telescopes is determined with a precision of about 17\% at low energies and 15\% at medium energies.
The systematic error of the slope of the energy spectra is estimated to be 0.15.
At medium energies the error in the flux normalization (without the energy scale uncertainty) is estimated to be 11\%. 
At low energies the systematic errors are in general larger and the flux normalization is known with a precision of about 19\%.
Additionally, we find that the Crab Nebula spectrum reconstructed by the MAGIC telescopes is consistent with other experiments within 20-30\%.

\begin{figure}[t]
\centering 
\includegraphics[width=0.49\textwidth]{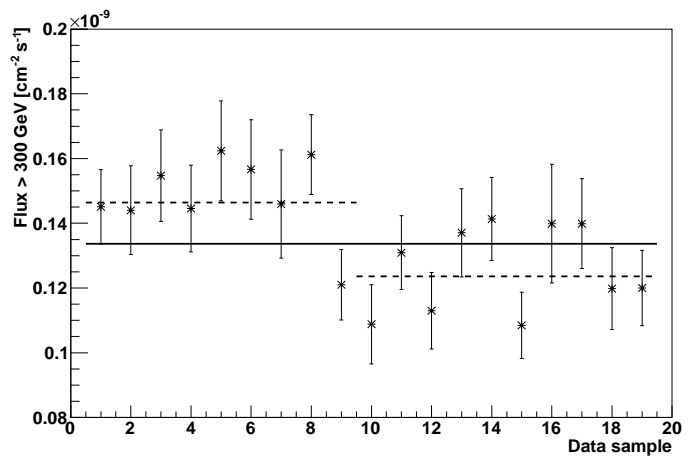}
\caption{
\komm{
Integrated flux above 300$\,$GeV of the Crab Nebula, as the function of the data sample number. 
The solid horizontal line shows the mean flux in the whole period. 
The dashed lines show the mean fluxes in November 2009 (left) and January 2010 (right). 
}
}\label{fig_lc}
\end{figure}

\komm{
\subsection{Run to run systematic uncertainty} 
The total systematic error estimated in the previous section can be used when the data from the MAGIC telescopes are compared with the data of other instruments. 
However, large part of the studied here sources of the systematic errors, will result in a (nearly) constant offset which will affect all the data in the same way. 
Therefore, the remaining, relative systematic error which may change from one data run to another will be certainly smaller. 
In order to investigate this effect we divided our data into 40 min sub-samples. 
Such time binning allows us to compute the integrated flux (above 300$\,$GeV) of the Crab Nebula with a precision of about 9\% (see Fig.\ref{fig_lc}).
}

\komm{
The standard deviation of the resulting distribution of fluxes computed from individual data runs is equal to 12\%.
While the fluxes both in November 2009 and January 2011 can be well fitted with a hyphotesis of a constant flux ($\chi^2 / n_{\mathrm dof} = 8.8/8$ and $10.1/9$ respectively), there is a clear shift of about $17\%$ between both months.
We conclude that the relative systematic error on short time scales (within a few days) with no drastically changing observational conditions is below 9\%. 
On the longer time scales (months) the relative systematic may result in variations of the estimated flux of the order of $\sim 17\%/\sqrt{2} = 12\%$, which is similar to the value obtained by the H.E.S.S telescopes \citep{ah06}.
Note, that since the Crab Nebula may show intrinsic variability, the values derived in this section should be treated as an upper limit on the run to run systematic uncertainty of the MAGIC telescopes.
}
%-----------------------------------------------------------------------------
\section{Conclusions}\label{sec:concl}
We evaluated the performance of the MAGIC telescopes using both MC simulations and a sample of 9$\,$h of Crab Nebula data. 
We obtained that the energy threshold of the MAGIC telescopes (defined as the peak of the energy distribution for a source with a -2.6 spectral index) is 50$\,$GeV. 
The upgrade of the MAGIC project with the second telescope led to a very good integral sensitivity of $(0.76\pm0.03)$\% of the Crab Nebula flux in 50$\,$h of the effective time in the medium energy range ($> 290\,$GeV).
An even bigger improvement (a factor of $\sim 3$  with respect to MAGIC~I single-telescope observations) is obtained at lower energies, thus allowing to reduce the needed observation time by a factor of $9$. 
This makes the MAGIC telescopes an excellent instrument for observations of gamma rays with energies around 100$\,$GeV.  
Thanks to the new Stereo DISP RF method, the angular resolution has improved as well ($\sim0.07^\circ$ at 300$\,$GeV).
The energy resolution is as good as 16\% at medium energies. 

We investigated different sources of systematic uncertainties and found that for a strong source like the Crab Nebula they dominate over the statistical errors. 
The spectrum of the Crab Nebula obtained with the MAGIC Stereo system is consistent with the spectra of other VHE Cherenkov telescopes.

\section*{Acknowledgements}
We would like to thank the Instituto de Astrof\'{\i}sica de
Canarias for the excellent working conditions at the
Observatorio del Roque de los Muchachos in La Palma.
The support of the German BMBF and MPG, the Italian INFN, 
the Swiss National Fund SNF, and the Spanish MICINN is 
gratefully acknowledged. This work was also supported by 
the Marie Curie program, by the CPAN CSD2007-00042 and MultiDark
CSD2009-00064 projects of the Spanish Consolider-Ingenio 2010
programme, by grant DO02-353 of the Bulgarian NSF, by grant 127740 of 
the Academy of Finland, by the YIP of the Helmholtz Gemeinschaft, 
by the DFG Cluster of Excellence ``Origin and Structure of the 
Universe'', by the DFG Collaborative Research Centers SFB823/C4 and SFB876/C3,
and by the Polish MNiSzW grant 745/N-HESS-MAGIC/2010/0.

\end{document}